\documentclass[11pt]{article}
\pdfoutput=1
\usepackage[utf8]{inputenc}
\usepackage{jheppub}
\usepackage{graphicx}
\usepackage{dcolumn}
\usepackage{bm}
\usepackage{color}
\usepackage[dvipsnames]{xcolor}
\usepackage{url}
\usepackage{float}
\usepackage{amsmath,amssymb,graphicx}
\usepackage{amsmath}
\usepackage{multirow}
\usepackage{amssymb}
\usepackage{mathrsfs}
\usepackage{comment}
\usepackage{accents}
\usepackage{epsfig}
\usepackage[lofdepth,lotdepth,caption=false]{subfig}
\usepackage{mathtools}
\usepackage{mathrsfs}
\usepackage{mathtools}
\usepackage{booktabs}
\usepackage{slashbox}
\usepackage{adjustbox}
\usepackage{lineno}
\usepackage{appendix}
\usepackage{hyperref}

\DeclarePairedDelimiterX\braket[2]{\langle}{\rangle}{#1 \delimsize\vert #2}

\newcommand{\beq}{\begin{equation}}
\newcommand{\eeq}{\end{equation}}
\newcommand{\beqa}{\begin{eqnarray}}
\newcommand{\eeqa}{\end{eqnarray}}
\newlength\figureheight 
\newlength\figurewidth 
\definecolor{MyGrey}{rgb}{0,0,0} 
\definecolor{MyDarkBlue}{rgb}{0.3,0.3,0.9} 
\definecolor{MyLightBlue}{rgb}{0.22,0.51,0.9}

\definecolor{MyRed}{rgb}{0.96,0.09,0.12}

\usepackage{cancel}

\def\beq{\begin{equation}}
\def\eeq{\end{equation}}

\title{On The Decaying-Sterile-Neutrino Solution to the Electron (Anti)Neutrino Appearance Anomalies}

\author{Andr\'e de Gouv\^ea$^{1}$,}

\author{O. L. G. Peres$^{2}$,}

\author{Suprabh Prakash$^{2}$}

\author{and G.  V. Stenico$^{2,3}$}

\affiliation{
$^{1}$Northwestern University, Department of Physics \& Astronomy, 2145 Sheridan Road, Evanston, IL 60208, USA \\
$^{2}$Instituto de F\'isica Gleb Wataghin - UNICAMP, 13083-859, Campinas, SP, Brazil \\
$^{3}$Fermi National Accelerator Laboratory P.O. Box 500, Batavia, Illinois 60510, USA}

\emailAdd{degouvea@northwestern.edu}
\emailAdd{orlando@ifi.unicamp.br}
\emailAdd{sprakash@ifi.unicamp.br}
\emailAdd{gstenico@ifi.unicamp.br}
\date{\today} 

\abstract{We explore the hypothesis that the unexplained data from Liquid Scintillator Neutrino Detector (LSND) and MiniBooNE experiments are evidence for a new, heavy neutrino mass-eigenstate that mixes with the muon-type neutrino and decays into an electron-type neutrino and a new, very light scalar particle. We consider two different decay scenarios, one with Majorana neutrinos, one with Dirac neutrinos; both fit the data equally well. We find a reasonable, albeit not excellent, fit to the data of MiniBooNE and LSND. The decaying-sterile-neutrino hypothesis, however, cleanly evades constraints from disappearance searches and precision measurements of leptonic meson decays, as long as $1~{\rm MeV}\gtrsim m_4\gtrsim 10$~keV. The Short-Baseline Neutrino Program  (SBN) at Fermilab should be able to definitively test the decaying-sterile-neutrino hypothesis. }
\begin{document}
\maketitle

\section{Introduction}\label{sec:intro}

Over the last several decades, a variety of revolutionary neutrino puzzles evolved into our current understanding of the neutrino sector of fundamental particle physics. A few of these puzzles, however, remain unresolved. Among them are data from the Liquid Scintillator Neutrino Detector (LSND) and MiniBooNE experiments.

The LSND collaboration looked for $\bar{\nu}_{e}$-candidate events at a detector located dozens of meters away from a stopped-pion target. Stopped $\pi^+$ decay into $\mu^+\nu_{\mu}$ and the muon subsequently decays, $\mu^+\to e^+\nu_{e}\bar{\nu}_\mu$, yielding a well-characterized flux of $\nu_{e},\nu_{\mu},\bar{\nu}_{\mu}$ and, most relevant, no $\bar{\nu}_e$. LSND observes a very significant excess -- more than 4~sigma -- of $\bar{\nu}_{e}$-candidate events  \cite{Aguilar:2001ty}. 

The MiniBooNE experiment was designed to test the oscillation-interpretation of the LSND data, discussed in more detail in the next paragraph. The detector was located downstream of a pion-decay-in-flight neutrino or antineutrino beam (mostly $\pi^+\to\mu^+\nu_{\mu}$ or $\pi^-\to\mu^-\bar{\nu}_{\mu}$). The experimental baseline $L$ was chosen such that, for typical neutrino energies $E_{\nu}$, the value of $L/E_{\nu}$ matched that of the LSND experiment. The MiniBooNE collaboration reported a combined 4.7~sigma excess of $\nu_e$- \cite{AguilarArevalo:2007it,Aguilar-Arevalo:2018gpe} and $\bar{\nu}_{e}$-candidate events  \cite{AguilarArevalo:2010wv} -- the detector has very limited charge-discrimination capabilities while running in both the neutrino-beam and antineutrino-beam modes. If both the LSND and MiniBooNE data are a consequence of the same unexplained phenomenon, the combined evidence is at the 6~sigma level \cite{Aguilar-Arevalo:2018gpe}.

Under the assumption that there are no unaccounted for ``mundane'' explanations to these two excesses -- unidentified background processes, problems with modelling the neutrino scattering process, detector-related effects, etc -- these so-called short-baseline anomalies\footnote{The short-baseline anomalies also include the reactor and gallium anomalies. For recent summaries of these data, see, for example, Refs. \cite{Collin:2016aqd,Dentler:2018sju,Boser:2019rta,Diaz:2019fwt}. We will have nothing to say about these here other than the fact that the hypothesis we will be investigating cannot account for either of them.} translate into new more physics -- on top of nonzero active neutrino masses -- in the neutrino sector. The simplest new-physics interpretation to the data from LSND and MiniBooNE is to postulate that a $\nu_{\mu}$ ($\bar{\nu}_{\mu}$) has nonzero probability of being detected as a $\nu_e$ ($\bar{\nu}_e$). Neutrino oscillations can lead to this phenomenon. In light of all other evidence for neutrino oscillations, the neutrino oscillation interpretation to the the short-baseline anomalies requires the introduction of a fourth neutrino mass eigenstate $\nu_4$ associated to a mass-squared difference $\Delta m^2\sim 1$~eV$^2$. The data point to new mixing parameters such that $|U_{e4}|^2|U_{\mu 4}|^2\sim 10^{-3}$ \cite{Aguilar-Arevalo:2018gpe}. In this scenario, the new flavor eigenstate is postulated to have no gauge quantum numbers and is hence dubbed a sterile neutrino. While this eV-scale sterile-neutrino hypothesis fits all data associated with searches for $\nu_{\mu}\to\nu_e$ appearance, it is in conflict with other data, including neutrino disappearance data at short-baselines. Very roughly, the reason for this is that there is no incontrovertible evidence for neutrino disappearance at short-baselines. These failed searches constrain $|U_{e4}|^2$ and $|U_{\mu4}|^2$ to be less than several percent and hence fail to satisfy $|U_{e4}|^2|U_{\mu 4}|^2\sim 10^{-3}$. More quantitatively, global fits to the world's neutrino data indicate that the eV-scale sterile-neutrino hypothesis is not a satisfactory explanation for the short-baseline anomalies. See, for example, Refs. \cite{Collin:2016aqd,Dentler:2018sju,Boser:2019rta,Diaz:2019fwt} for recent analyses and discussions. 

Here, we revisit a different solution to the LSND and MiniBooNE puzzle. Instead of assuming that a fourth eV-scale neutrino is produced coherently during pion or muon decay, we postulate that a heavier fourth neutrino mass eigenstate is produced in the neutrino source and that this new neutrino state decays into an electron-type neutrino and a new, effectively massless scalar particle \cite{PalomaresRuiz:2005vf}. The decay is prompt enough such that, a significant portion of the time the daughter neutrino can interact in the detector and lead to an excess of $\nu_e$- and $\bar{\nu}_{e}$-candidate events. This hypothesis was first raised to explain the LSND results \cite{PalomaresRuiz:2005vf}. Radiative sterile-neutrino decays were also explored as a potential explanation to the observations reported by LSND and  MiniBooNE~\cite{Gninenko:2009ks,Gninenko:2009yf,Gninenko:2010pr,Dib:2011jh}. We do not consider these here. 

We extend the analysis in Ref.~\cite{PalomaresRuiz:2005vf} to include the most recent data from the MiniBooNE experiment, and ask whether the decaying-sterile-neutrino hypothesis is a good fit to the data. We explore different decay scenarios with Majorana neutrinos and Dirac neutrinos. These are spelled out in  Section~\ref{sec:form}. We also explore how well the decaying-sterile-neutrino hypothesis will be tested by the Short-Baseline Neutrino Program (SBN) at Fermilab. Details and results, along with a description of how we treat the data from LSND and MiniBooNE, are discussed in Section~\ref{sec:sim}. A short summary of our findings is presented in Section~\ref{sec:conc}.

\section{Formalism}\label{sec:form}

We postulate the existence of a fourth neutrino mass eigenstate. Since we want to explain the data from LSND and MiniBooNE, the fourth neutrino must have a nonzero $\nu_{\mu}$ component. We don't need a nonzero $\nu_{\tau}$ or $\nu_e$ component so we set these to zero. A very small $\nu_e$ or $\nu_{\tau}$ component would not modify our results in a significant way. In other words, $U_{e4}=U_{\tau4}=0$ and $\nu_\mu = U_{\mu i}\nu_i$, $i=1,2,3,4$, $U_{\mu4}\neq 0$.

We further introduce a new interaction that allows $\nu_4$, with mass $m_4$, to decay into a new, very light scalar field $\phi$ and a $\nu_e$. There are different effective Lagrangians capable of mediating this phenomenon \cite{Schechter:1981cv,PhysRevLett.45.1926,Berezhiani:1990sy,1988SvA....32..127D,GELMINI1981411,Gelmini:1983ea,BERTOLINI1988714,SANTAMARIA1987423,Cogollo:2008zc}. Here, we concentrate on two possibilities.

If the neutrinos are Dirac fermions, there are several distinct ways of coupling a heavy neutrino to a light neutrino and a scalar field $\phi$. These are associated with the transformation properties of $\phi$ under lepton-number (or lepton-number-minus-baryon-number) and the parity-violating properties of the new interaction. We will assume that the new scalar field $\phi$ is a standard model gauge singlet and that it carries zero lepton number. We will also assume that the new interaction violates parity maximally and, like the weak-interactions, only couples to left-chiral light neutrinos. Since we are interested in the decay to $\nu_e$, at low-energies, the interaction that mediates the heavy neutrino decay is
\begin{equation}
{\cal L}_{\rm Dirac} = - g_{\rm D} \nu^c_4\nu_e \phi + H.c.~,
\label{eq:dirac}
\end{equation}
where, to facilitate comparisons to the Majorana case, we express the neutrino fields as two-component Weyl fermions\footnote{Using 4-component Dirac spinors, ${\cal L}_{\rm Dirac} = - g_{\rm D} \bar{\nu}_4\frac{(1-\gamma_5)}{2}\nu_e \phi + H.c.$.}. $\nu_e$ is the field associated with the left-chiral $\nu_e$ and the $\nu_4^c$ is the field associated with the left-chiral $\bar{\nu}_4$. Note that $\nu_e$ is a linear superposition of the light neutrino mass eigenstates, $\nu_e=U_{ei}\nu_i$, $i=1,2,3$. We choose this specific decay in order to maximize the effect of the sterile-neutrino decay at MiniBooNE and LSND and in order to minimize the effect at experiments sensitive to $\nu_{\mu}$ or $\nu_{\tau}$ in the final state.  Note that in the Lagrangian in Eq.~(\ref{eq:dirac}), we do not have a $\nu_{\mu}$ or a $\nu_{\tau}$ in the final state from the $\nu_4$ decay. We will be interested in  $L/E_{\nu}$  values such that ordinary neutrino oscillations, driven by $m_2^2-m_1^2$ and $m_3^2-m_1^2$, do not have time to modify neutrino flavor evolution. We will also be interested in $m_4$ values that are much larger than $m_{1,2,3}$ and will treat the $\nu_e$  as a massless particle. This means that all daughters of the $\nu_4$ decay mediated by Eq.~(\ref{eq:dirac}) are left-handed $\nu_e$ while all the daughters of the $\bar{\nu}_4$ decay mediated by Eq.~(\ref{eq:dirac}) are right-handed $\bar{\nu}_e$. Other choices lead to different final states. For example, one can choose $\phi$ to carry lepton number two in such a way that the decay process is $\nu_4\to \bar{\nu}_e\phi$, or one can choose an effective Lagrangian proportional to $\nu_4\nu^c_e$ so that all $\nu_e$ produced in the decay of $\nu_4$ are right-handed. The choice above -- Eq.~(\ref{eq:dirac}) -- maximizes the ``visibility'' of the daughter $\nu_e$. In the limit where the light neutrino masses are negligible -- an excellent approximation here -- the left-handed $\nu_e$ is perfectly aligned with the left-chiral interaction field, while the right-handed $\nu_e$ is perfectly sterile as far as the weak interactions are concerned. For a more detailed, recent discussion of these issues, see, for example, Ref. \cite{deGouvea:2019goq}. Note that it is easy to express Eq.~(\ref{eq:dirac}) in a way that explicitly preserves the $SU(2)\times U(1)$ gauge symmetry of the standard model: $\nu^c_4\nu_e \phi \to  \nu^c_s(L_eH) \phi/\Lambda$, where $L_e$ is the electron-flavor lepton-doublet, $H$ is the Higgs boson field, and $\Lambda$ is the effective scale of the physics that leads to the decay Lagrangian.

If the neutrinos are Majorana fermions and one only adds one new Weyl fermion to the low-energy particle content of the standard model -- $\nu_4$ -- along with the gauge-singlet scalar field $\phi$, the Lagrangian that mediates tree-level $\nu_4$ decay at low energies is 
 \begin{equation}
{\cal L}_{\rm Majorana} = - g_{\rm M} \nu_4\nu_e \phi + H.c.~.
\label{eq:maj}
\end{equation}
Here it is not meaningful to assign lepton-number charge to the $\phi$-field. Since we are interested in the limit where the light neutrino masses are negligible, it is meaningful and convenient to talk about $\nu_e$ -- always left-handed -- and $\bar{\nu}_e$ -- always right-handed. In this case, Eq.~(\ref{eq:maj}) mediates both $\nu_4\to \nu_e\phi$ and $\nu_4\to \bar{\nu}_e\phi$, both with the same branching ratio at the tree-level. Here it is also easy to express Eq.~(\ref{eq:maj}) in a way that explicitly preserves the $SU(2)\times U(1)$ gauge symmetry of the standard model:  in the limit $U_{\mu4}\ll 1$, $\nu_4\nu_e \phi \to  \nu_s(L_eH) \phi/\Lambda$, where $\nu_s$ is the left-handed sterile neutrino field and $\Lambda$ is the effective scale of the physics that leads to the decay Lagrangian\footnote{Of course, one can also ultraviolet-complete interactions among the active neutrinos: $\nu_{\alpha}\nu_{\beta}\phi~\to~ (L_{\alpha}H)(L_{\beta}H)\phi$ and use a judicious combination of this and the sterile interaction and perfectly realize Eq.~(\ref{eq:maj}) at low energies.}. Again, for a more detailed, recent discussion of these issues, see, for example,~\cite{deGouvea:2019goq}.

Henceforth, we will treat $\nu_e$ and $\phi$ as massless particles. For Dirac neutrinos, in the ultra-relativistic approximation ($\beta_4  \rightarrow 1$), the differential decay rate of a $\nu_4$ in the laboratory reference frame with helicity $r$ and energy $E_{4}$ into a $\nu_e$ with helicity $s$ and energy $E_{e}$ is \cite{PalomaresRuiz:2005vf}
\beq\label{dkrate}
\frac{d\Gamma_{\nu_4^r\to \nu_e^s}(E_{4})}{dE_{e}} = 
\frac{1}{16 \pi E_{4}^2} \, |\mathcal{M}_{rs}|^2,
\eeq
and the matrix element is 
\beq\label{matrixel}
|\mathcal{M}_{rs}|^2 = |g_{\rm D}|^2 \, m_4^2 \times
\left\{
\begin{array}{l@{\quad}l}
E_{e}/E_{4} & r=s \\
(1 - E_{e}/E_{4}) & r \neq s 
\end{array} \right. \,.
\eeq
The same expression, of course, holds for the decay of $\bar{\nu}_4$. In the scenario of interest (Eq.~\ref{eq:dirac}), all daughter $\nu_e$ are left-handed and all $\nu_4$ are produced via the weak interactions and are relativistic in the laboratory reference frame. We are also interested in $\nu_4$ masses that are much smaller than the mass of the muon. Therefore, in the laboratory reference frame, virtually all $\nu_4$ are left-handed and the energy spectrum of the daughter neutrinos is proportional to  $E_{e}/E_{4}$. Here, we are interested in $\nu_4$ masses below a few MeV and neutrino energies characteristic of the MiniBooNE and LSND experiments. This means the decay products are emitted in the forward direction and inherit the angular distributions of the parent. The properties of decay-products were discussed in detail in Ref.~\cite{Lindner:2001fx}.

The total decay width for $\nu_4\to \nu_e + \phi$ in the laboratory reference frame is
\beq\label{dk}
\Gamma_{4e} = 
\frac{|g_{\rm D}|^2 \, m_4^2}{32 \pi E_{4}} \, .
\eeq

The situation is very similar for Majorana neutrinos. Eq.~(\ref{dkrate}) holds along with Eq.~(\ref{matrixel}) as long as one replaces $g_{\rm D}\to g_{\rm M}$ and allows for both $\nu_4\to \nu_e\phi$ and $\nu_4\to \bar{\nu}_e\phi$, keeping in mind that the $\nu_e$ are all left-handed and the $\bar{\nu}_e$ are all right-handed. Here, the $\nu_4$ are produced via the weak interactions, are relativistic in the laboratory reference frame, and are much lighter than the muon\footnote{This also implies that it is meaningful to talk about $\nu_4$ and $\bar{\nu}_4$, even if neutrinos are Majorana fermions.}. Hence, in the lab frame, we expect the energy spectrum of the daughter neutrinos to be proportional to $E_{e}/E_{4}$ (harder) while that of the daughter antineutrinos to be proportional to $(1-E_{e}/E_{4})$ (softer).

The total decay rate of $\nu_4$ in the laboratory reference frame is given by Eq.~(\ref{dk}) with $g_{\rm D}\to g_{\rm M}$ and an overall factor of 2~\cite{Kim:1990km}, accounting for the fact that there are two different allowed decay modes:
\beq\label{dk_maj}
\Gamma_{4e} = 
\frac{|g_{\rm M}|^2 \, m_4^2}{16 \pi E_{4}} \, .
\eeq

It is straight-forward to compute, for a neutrino produced in a charged-current process involving muons, the energy and flavor of the neutrinos that reach the detector. In our computations, we make use of the results in \cite{PalomaresRuiz:2005vf}, to which we refer to in more details, adapting the relevant expressions for the decay-scenarios of interest. For a recent, more complete treatment, that combines oscillation and decay effects, see, for example, Ref.~\cite{Lindner:2001fx}. Here, instead, we summarize the qualitative impact of $\nu_4$ production and decay. This discussion will help inform the results we present in the following sections. 

We are interested in $m_4\gg m_{1,2,3}$ and, in a charged-current process involving muons, the $\nu_4$ is produced incoherently relative to $\nu_{1,2,3}$. Hence, when, for example, a pion decays into a muon and a neutrino, the neutrino is either a $\nu_4$, with probability $|U_{\mu4}|^2$, or the orthogonal state\footnote{A ``light'' $\nu_{\mu}$, proportional to $U_{\mu i}\nu_i$, $i=1,2,3$.}, with probability $1-|U_{\mu4}|^2$. If the initial state is a $\nu_4$, it will reach the detector with probability $e^{-\Gamma_{4e} L}$, where $L$ is the baseline and $\Gamma_{4e}$ is the $\nu_4$ decay width, see Eq.~(\ref{dk}) or (\ref{dk_maj}). Hence, the probability that the neutrino will behave like a $\nu_{\mu}$ in the detector is
\beq
P_{\mu\mu} = (1-|U_{\mu4}|^2)^2 + (|U_{\mu 4}|^2)^2e^{-\Gamma_{4e} L}.
\label{eq:pmumu}
\eeq
In the limit where $\nu_4$ is very long-lived, $\Gamma_{4e} L\ll 1$, $P_{\mu\mu}=1-2|U_{\mu 4}|^2(1-|U_{\mu4}|^2)$. This agrees with the $\nu_{\mu}$ survival probability assuming there is a stable $\nu_4$ and it is produced incoherently or, equivalently for the purposes of this setup, the new mass-squared difference is very large, $\Delta m^2_{41}L/E\gg 1$, where  $\Delta m^2_{41}\equiv m_4^2-m_1^2$, and the oscillations average out. Instead, in the limit where the decay is fast $\Gamma_{4e} L\gg 1$, $P_{\mu\mu} = (1-|U_{\mu4}|^2)^2$.

The parent particle will yield a $\nu_e$ in the final state only if $\nu_4$ decays because $U_{e4}\equiv 0$ and both $\nu_4$ and the state proportional to $U_{\mu i}\nu_i$, $i=1,2,3$, are orthogonal to $\nu_e$. If the $\nu_4$ decays before reaching the detector -- this happens with probability $(1-e^{-\Gamma_{4e} L})$ -- a $\nu_e$ or a $\bar{\nu}_e$ with some energy less than the original parent energy will arrive at the detector\footnote{In the laboratory frame, the angular distribution of the decay is very forward peaked since the $\nu_4$ are ultra-relativistic. Hence, we assume all daughter-neutrinos reach the detector.} with probability $B_e$ or $B_{\overline{e}}$. In the Dirac case of interest here, $B_e=1$, $B_{\overline{e}}=0$, while in the Majorana case $B_e=B_{\bar{e}}=0.5$. The probability that the $\nu_e$ or $\bar{\nu}_e$ emerges with energy $E_e$ is proportional to Eq.~(\ref{dkrate}). The same happens for $\bar{\nu}_4$ decays. In summary,
\beq
P_{\mu e} = P_{\overline{\mu}\, \overline{e}}\propto B_e|U_{\mu4}|^2(1-e^{-\Gamma_{4e} L}),~~~ P_{\mu \overline{e}} = P_{\overline{\mu}\, e}\propto B_{\overline{e}}|U_{\mu4}|^2(1-e^{-\Gamma_{4e} L}),
\label{eq:pmue}
\eeq
and the same-helicity (opposite-helicity) final state has a harder (softer) spectrum. Note that, strictly speaking, $P_{\mu e} (P_{\overline{\mu}\,\overline{e}})$  and $P_{\mu \bar{e}} (P_{\overline{\mu}\, e})$ are not probabilities. 

Qualitatively, it is easy to see why this hypothesis can outperform the standard (3+1)-oscillation hypothesis  ~\cite{Collin:2016aqd,Dentler:2018sju,Boser:2019rta,Diaz:2019fwt}. In the (3+1)-oscillation scenario, $P_{\mu e}\propto |U_{\mu4}|^2|U_{e4}|^2$ while the survival probabilities of $\nu_{\mu}$ and $\nu_e$ are, respectively, $1- P_{\mu\mu}\propto |U_{\mu4}|^2(1-|U_{\mu4}|^2)$  and $1- P_{ee}\propto |U_{e4}|^2(1-|U_{e4}|^2)$. A sizable $P_{\mu e}$ requires both a non-negligible $|U_{\mu4}|^2$ and $|U_{e4}|^2$ which, in turn, are constrained by disappearance searches~\cite{Peres:2000ic}. In the sterile-decay scenario, the original electron neutrino does not change  and, $1- P_{\mu\mu}\propto |U_{\mu4}|^2(1-|U_{\mu4}|^2)$, similar to the oscillation scenario, especially in the limit of small $|U_{\mu4}|^2$. Instead, the role of $|U_{e4}|^2$ is played by $B_e(1-e^{-\Gamma_{4e} L})$. $\Gamma_{4e}$ is not constrained by $\nu_e$-disappearance. Instead, it is constrained by non-oscillation experiments, as we quickly summarize in the next subsection, and we find that reasonably large values of $\Gamma_{4e} L$ are allowed for the $L/E_{\nu}$ values of interest. In the case of Majorana neutrinos, one half of the neutrinos will decay into antineutrinos, and vice-versa. This means that, in the case of the LSND experiment, some of the $\bar{\nu}_e$-excess events arises from parent $\nu_{\mu}$ created in the decay of the stopped $\pi^+$, while half of the decaying-component associated with the $\bar{\nu}_{\mu}$ from the Michel decay will behave like a $\nu_e$ and will not contribute to the $\bar{\nu}_e$-excess. In the case of MiniBooNE, the excess of $\nu_e$ and $\bar{\nu}_e$ events will be associated to both $\nu_{\mu}$ and $\bar{\nu}_{\mu}$ parents. Since the wrong-sign contamination is different between neutrino-mode running and  antineutrino-mode running, we expect the excesses observed in the case of the neutrino and antineutrino beams to be slightly different. We return to these issues in the discussion of our results, in Sec.~\ref{sec:sim}.

\subsection{Constraints on New Neutrinos and Neutrino--Scalar Interactions}\label{sub:2}

There are several bounds on the new-physics parameters we are introducing here: $m_4$, $g_{\rm D,M}~\equiv~g$, and $|U_{\mu 4}|$. We will discuss oscillation-related bounds in the next sections and here we summarize non-oscillation results. 

Searches for neutral heavy leptons constrain $|U_{\mu 4}|^2$ as a function of $m_4$. Keeping in mind that we are interested in constrains assuming $\nu_4$ decays, as far as non-neutrino-oscillation experiments are concerned,  invisibly, $|U_{\mu 4}|^2\lesssim 10^{-2}$ for $m_4\gtrsim 1$~MeV (see Refs. \cite{deGouvea:2015euy,Bryman:2019bjg} for recent quantitative analyses). The bounds are significantly weaker for smaller values of the $m_4$. For $m_4\simeq 1$~MeV, the strongest bounds come from precision measurements of $\pi\to\mu\nu$. Bounds from $\nu_{\mu}$ disappearance, as we will discuss later, are around $|U_{\mu 4}|^2\lesssim 10^{-2}$ for $m_4\gtrsim 10$~eV and hence will dominate for $m_4\lesssim 1$~MeV.

The couplings $g$ of neutrinos to other neutrinos and a scalar particle, in the region of parameter space of interest here, are also best constrained by leptonic meson decays, especially the decays of pions and kaons (e.g. $K\to \mu\nu\phi$). The bound on $g$ depends on both the nature of the decay and on $|U_{\mu 4}|^2$. Here, conservatively, we use the results from Ref.~\cite{Pasquini:2015fjv}, which translate into
\beq
g^2 |U_{\mu 4}|^2 < 1.9 \times 10^{-7}. 
\label{bound2}
\eeq

As far as short-baseline experiments, we are sensitive to $|U_{\mu 4}|^2$ and $\Gamma_{4e}\propto (gm_4)^2$, see Eqs.~(\ref{dk}, \ref{dk_maj}). As will be discussed in great detail in the next couple of sections, we will be interested in $(gm_4)^2 |U_{\mu 4}|^2 \sim 1$eV$^2$ or 
\beq
g^2 |U_{\mu 4}|^2 \sim \left(\frac{1~{\rm eV}}{m_4}\right)^2,
\eeq
so the constrain in Eq.~(\ref{bound2}) can be easily satisfied for $m_4 \gtrsim 10$~keV.

In summary, for $1~{\rm MeV}\gtrsim m_4 \gtrsim 10~{\rm keV}$, we expect to avoid all non-oscillation bounds with relative ease. We return to these in Sec.~\ref{sec:sim}.

\section{Simulations and Results}\label{sec:sim}

Here we provide details of the data we analyse and discuss how well they fit the decaying-sterile-neutrino hypothesis. We also discuss the details of our simulation of data from the SBN program and how sensitive it is to the decaying-sterile-neutrino hypothesis.

\subsection{LSND}\label{sub:lsnd}
The LSND experiment~\cite{Athanassopoulos:1996ds} ran at the Los Alamos Neutron Science Center (LASCE) from 1993 to 1998. The experiment was designed to look for $\bar{\nu}_{e}$ from a pion-decay-at-rest neutrino source~\cite{Aguilar:2001ty}. LSND consisted of a cylindrical tank filled with 167 tons of mineral oil doped with a low concentration of liquid scintillator. This combination allows the detection of both Cherenkov and scintillation light, which are collected by 1220 photo-multiplier tubes (PMT) that surround the detector inner wall. Neutrinos are produced by the interaction of a 798~MeV proton beam with a production target, where positive pions stop at the beam dump and decay at rest into positive muons $\left( \pi^{+} \rightarrow \mu^{+} + \nu_{\mu} \right)$. The distance between the beam dump and the longitudinal center of LSND is 30 meters. The positive muons also decay at rest $\left( \mu^{+} \rightarrow e^{+} + \nu_{e} + \bar{\nu}_{\mu} \right)$. The Michel $\bar{\nu}_{\mu}$ would lead to a $\bar{\nu}_{e}$ signal in the presence of neutrino oscillations or other flavor-changing mechanism. The  $\bar{\nu}_{e}$ are detected via inverse beta decay (IBD), $\bar{\nu}_{e} + p \rightarrow n + e^{+}$, where the positron leads to Cherenkov and scintillation light inside mineral oil. The outgoing neutron manifests itself as subsequent scintillation light as it is captured on proton and a 2.2~MeV photon is emitted~\cite{Conrad:2016sve}. LSND makes use of this two-component signal to select a $\bar{\nu}_{e}$-candidate event sample.

In order to generate expected event rates for the different decay scenarios and fit them to the available data, we make use of the GLoBES \cite{Huber:2004ka, Huber:2007ji} c-library. Decay-at-rest fluxes were obtained from Ref.~\cite{Aguilar:2001ty}, and we use the IBD cross-section from Ref.~\cite{Strumia:2003zx}. In the case of  Majorana neutrinos, we expect $\bar{\nu}_{e}$ appearance from not only the $\bar{\nu}_{\mu}$ but also from the $\nu_{\mu}$ parents from $\pi^+$decay, as discussed in the previous section. We considered events associated to neutrino energies between 20 and 60 MeV. Finally, a Gaussian energy smearing with $\sigma(E_{\nu}) = 17\%/E_{\nu} [\rm{MeV}]$ was implemented to take into account the energy resolution of the experiment.

\begin{figure}[t]
\centering
\includegraphics[scale=0.6]{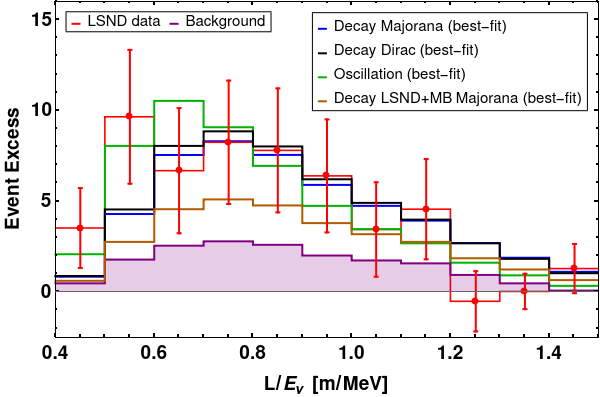}
\caption{Best-fit $\bar{\nu}_e$ spectra at LSND as a function of $L/E_{\nu}$ for the oscillation hypothesis and for the different decaying-sterile-neutrino scenarios discussed here. The data points and the background spectrum are from the LSND collaboration report, presented in Ref.~\cite{Aguilar:2001ty}.}
\label{fig:LSND-spec}
\end{figure}

We perform a $\chi^{2}$-analysis, including an overall normalisation error of 25\% for signal and background. Uncertainties in the neutrino flux, cross-section and efficiency lead to systematic errors between 10\% and 50\%, as discussed in Ref. \cite{Aguilar:2001ty}. The LSND background sources come mainly from intrinsic beam $\bar{\nu}_{e}$ and $\bar{\nu}_{\mu}$ events and are summarized in Table VIII of Ref.~\cite{Aguilar:2001ty}. In order to validate our analysis procedure, we first fit the two-flavor oscillation hypothesis and compare our results with those presented by the LSND collaboration \cite{Aguilar:2001ty}. When generating events, we introduce a normalization factor that allows us to mimic the total rates of the best-fit spectrum obtained by the LSND collaboration (Figure 24 of Ref.~\cite{Aguilar:2001ty}). Our best-fit oscillation spectrum (green histogram), in 11  bins of $L/E_{\nu}$, is depicted in Fig.~\ref{fig:LSND-spec}, along with the data  and backgrounds published by the collaboration; the best-fit point for the oscillation analysis is $\left(\sin^2 2\theta,\ \Delta m^{2}\right) = \left(0.0063,\ 7.2\ \rm{eV}^{2}\right)$ and the minimum value of $\chi^2$ is $\chi^2_{\rm min}=10.19$. Given the eleven bins we included in our analysis (and hence nine degrees of freedom), we conclude that two-flavor-oscillations are a good fit to the LSND data, as expected. The allowed regions of the $(\sin^22\theta,\ \Delta m^2)$ parameter space match well with those published by the LSND collaboration. With this agreement, we are confident we are capable of faithfully reproducing the data-analysis of LSND well enough to repeat the procedure for the decaying-sterile-neutrino hypothesis.

We generate neutrino event spectra for each set of decay parameters $\left( | U_{\mu 4} |^{2}, \ gm_{4} \right)$ and attempt to fit them to the LSND data, using a $\chi^{2}$-fit. The best-fit spectra in the case of Dirac and Majorana neutrinos are depicted, respectively, in black and blue in Fig.~\ref{fig:LSND-spec}. The results for the two hypotheses are very similar. The Majorana and Dirac cases are, in practice, identical, except for the fact that $B_{e}=1$ in the Dirac case and $B_{e}=0.5$ in the Majorana case. In the Majorana case, there is an antineutrino signal from $\nu_4\to \bar{\nu}_e\phi$ decays, but these are too low-energy and do not contribute significantly to the number of events. Since the effect of the decay is proportional to $|U_{\mu4}|^2B_e$, one can compensate for the change in $B_e$ by changing $|U_{\mu 4}|^2$ by a factor of two. The $\nu_4$ produced in DAR are monochromatic, with energy around 30~MeV. Hence, the $\bar{\nu}_e$ produced in $\nu_4\to \bar{\nu}_e\phi$ have very low energies and only populate the highest $L/E_{\nu}$-bins. The situation is made worse by the fact that the energy spectrum of the daughter $\bar{\nu}_e$ from the neutrino decay is soft, peaking (linearly) at zero energy. The overall result is that most $\bar{\nu}_e$ from $\nu_4\to \bar{\nu}_e\phi$ have too low energy to significantly contribute to the LSND excess. 

\begin{figure}[t]
\centering
\includegraphics[scale=0.35]{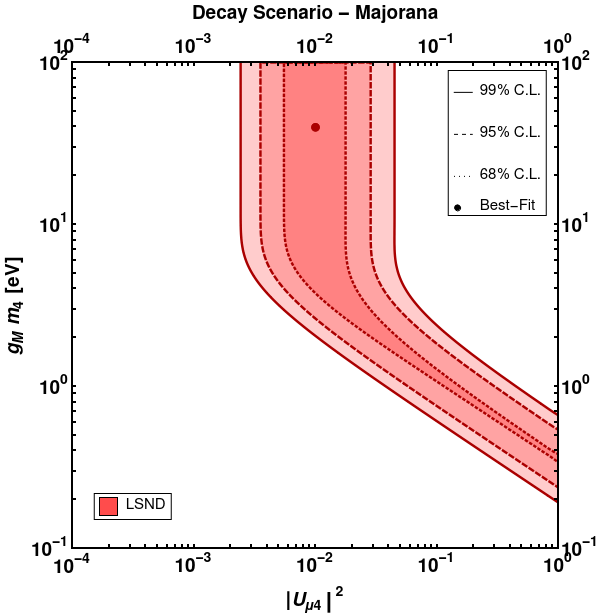}
\includegraphics[scale=0.35]{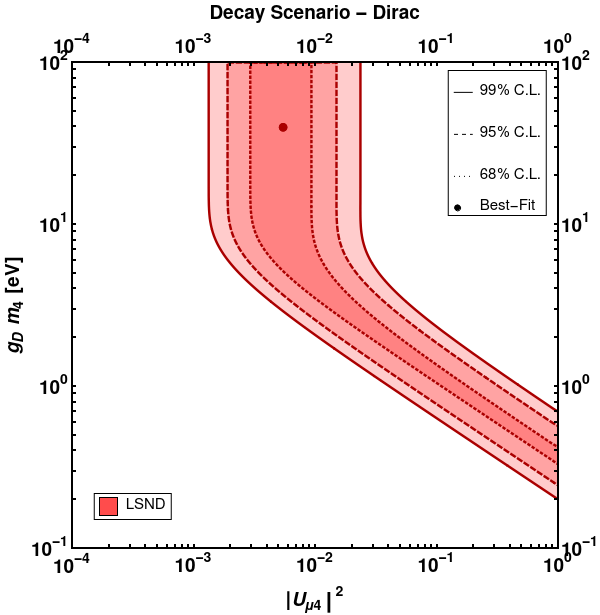}
\caption{Allowed regions of the $(|U_{\mu4}|^2,gm_4)$ parameter space when the decaying-sterile-neutrino hypothesis is matched against the LSND data assuming Majorana (left) or Dirac neutrinos (right). The dots indicate the best-fit-point and the lines represent the 99\% (solid), 95\% (dashed) and 68\% (dotted) confidence level (C.L.) curves.}
\label{fig:LSND-chisq}
\end{figure}

The best fit point falls in the region where the decay is fast so that, to zeroth order, all $\nu_4$ decay between production and detection. We estimate the goodness-of-fit by comparing $\chi^2_{\rm min}$=19.53 (20.17) in the Dirac (Majorana) cases with nine degrees of freedom and conclude the fit is acceptable (p-value around two percent). The quality of this fit is worse than that of the oscillation fit. This is due to fact that the energy spectrum of the daughter $\bar{\nu}_e$ is distorted towards lower energies compared with the energy spectrum of the parent $\bar{\nu}_4$. The allowed regions of the parameter space, along with the best-fit points, are depicted in Fig.~\ref{fig:LSND-chisq}.  Solid, dashed and dotted lines represent, respectively, the 99\%, 95\% and 68\% C.L. curves. As advertised, the results of the two decay scenarios are similar once one rescales the value of $|U_{\mu4}|^2$ by a factor of 2. 

\subsection{MiniBooNE}\label{sub:miniboone}

The  MiniBooNE experiment was designed to test the oscillation interpretation of the LSND data~\cite{AguilarArevalo:2008qa}. It consisted of a spherical tank filled with 800 tons of mineral oil and internally covered with 1280 PMTs to collect, mostly, Cherenkov light. The MiniBooNE detector is located 540 meters downstream from the neutrino source. In order to generate a neutrino flux, the booster neutrino beam (BNB), located at Fermilab, delivers 8.89 GeV protons that interact with a beryllium target. Charged mesons, like pions and kaons, are then produced and decay predominantly into muon neutrinos and antineutrinos. A magnetic focusing horn was used to sign-select the charged mesons, allowing, depending on the polarity of the horn, two neutrino-beam configurations: 1) \emph{neutrino mode}: positively-charged mesons are focused to create a high-intensity flux of neutrinos; 2) \emph{antineutrino mode}: negatively-charged mesons are focused to create a high-intensity flux of antineutrinos. MiniBooNE measures both $\nu_e$ and $\nu_{\mu}$, plus their antiparticles, and is sensitive to  $\nu_e \ \rm{and} \ \bar{\nu}_{e}$ appearance and $\nu_{\mu} \ \rm{and} \ \bar{\nu}_{\mu}$ disappearance. $\nu_{\mu, e}$ and $\bar{\nu}_{\mu,e}$ are identified as they scatter through the charged-current quasielastic (CCQE) process, yielding $\mu^{\pm},e^{\pm}$, respectively. These particles emit Cherenkov and scintillation light inside the detector, and muon-candidates are distinguished well from electron-candidates. 

We analyse MiniBooNE appearance data collected when the neutrino-beam was running in both neutrino and antineutrino modes~\cite{Aguilar-Arevalo:2018gpe, mbdata1}.  The MiniBooNE data set corresponds to $12.84\times10^{20}$ protons on target (POT) in neutrino mode and $11.27\times10^{20}$ POT in the antineutrino mode. We analyse the different data sets separately and combined.

We simulate MiniBooNE events in GloBES, where the CCQE cross-section information is available. Flux information was obtained from Ref.~\cite{AguilarArevalo:2008yp} and we include a Gaussian energy smearing function with $\sigma(E_{\nu}) = 30\%/\sqrt{E_{\nu} [\rm{GeV}]}$ to mimic the detector energy resolution. For the electron-like events, the analysis is done in the neutrino energy range $E_{\nu}\in \left[0.2, 3.0\right]~\rm{GeV}$ and the signal detection efficiencies for electron-like events are taken from \cite{mbdata2}. Background events are summarized in Table 1 and Figure 1 of Ref.~\cite{Aguilar-Arevalo:2018gpe}. Neutral current events are, strictly speaking, impacted by the $\nu_4$ decay, but the effect is negligible in the region of the parameter space in which we are interested. Changes to the neutral current (NC) event rate in this scenario are proportional to the maximum muon neutrino to sterile neutrino transition probability,  $\left(P_{\mu s}\right)^{\rm max} \leq 1-P_{\mu\mu}-P_{\mu e}-P_{\mu \bar{e}}\sim |U_{\mu4}|^2 (1-|U_{\mu4}|^2) $, using Eq.~(\ref{eq:pmumu}) and (\ref{eq:pmue}). This is small when $|U_{\mu4}|^2$ or $1-|U_{\mu4}|^2$ is small which, as we discuss in subsequent sections, is constrained to be small. Hence, we do not include decay effects in the background events. 

In our $\chi^{2}$ analysis that includes 11 bins from 0.2 to 3.0 GeV for neutrinos and anti-neutrinos, we take statistical and systematic errors into account by using the official MiniBooNE covariance matrices, available in Ref.~\cite{mbdata1}. These include correlations among $\nu_{e} \ (\bar{\nu}_{e})$ signal and background events and $\nu_{\mu} \ (\bar{\nu}_{\mu})$ events for the neutrino (antineutrino) mode. In the combined analysis, the correlations among all neutrino and antineutrino samples are considered. Here, we are ultimately interested in the region of the parameter space where the impact of the new physics on $\nu_{\mu}$-disappearance is very small, thanks to strong bounds from other experiments, discussed in Sec.~(\ref{sub:minos}). Hence, the only impact of the $\nu_{\mu}$ part of the data is to provide information concerning the neutrino flux and the neutrino scattering parameters. In other words, we are interested in gauging the impact of fitting the $\nu_e$ and $\bar{\nu}_e$ appearance data assuming the same new physics does not impact the $\nu_{\mu}$ and $\bar{\nu}_{\mu}$ data. In order to achieve this, we followed the prescription, discussed in Appendix E.4 of Ref.~\cite{Kopp:2013vaa}, of considering only the contribution of electron neutrino and antineutrino events (signal and background) in the fit, along with an extra component related to the uncertainty in the overall normalization of the spectrum. More details of the MiniBooNE analysis are available in Appendix~\ref{app:a}. 
We will use the minimum value of the $\chi^{2}$ in order to gauge the goodness-of-fit, using the 11 bins to compute the number of degrees of freedom. 

\begin{figure}[ht]
\centering
\includegraphics[scale=0.6]{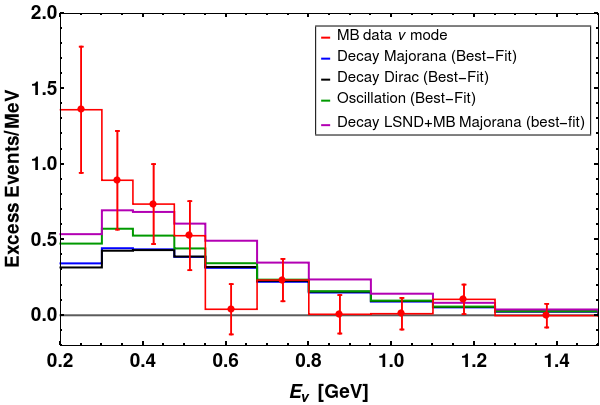}
\caption{Best-fit $\nu_e$ spectra at MiniBooNE, neutrino-mode, as a function of $E_{\nu}$ for the oscillation hypothesis and for the different decaying-sterile-neutrino scenarios discussed here. The data points are from the MiniBooNE collaboration report, presented in Ref.~\cite{Aguilar-Arevalo:2018gpe}. The last bin corresponding to $[1.5, 3.0]$ GeV is not shown here.} 
\label{fig:MB-spec}
\end{figure}

As in the LSND case, we first fit the MiniBooNE neutrino-mode and antineutrino-mode data with the two-flavor oscillation hypothesis. For the neutrino-mode data, our best-fit oscillation spectrum (green histogram), in bins of $E_{\nu}$, is depicted in Fig.~\ref{fig:MB-spec}, along with the excess data published by the collaboration; the best-fit point for the oscillation analysis is  $\left(\sin^2 2\theta,\ \Delta m^{2}\right) = \left(0.83,\ 0.036 \ \rm{eV}^{2}\right)$ and the minimum value of $\chi^2$ is $\chi^2_{\rm min}=9.46$. Given the eleven bins we included in our analysis (and hence nine degrees of freedom), we conclude that two-flavor-oscillations are a good fit to the MiniBooNE neutrino data, as expected. The allowed regions of the $(\sin^22\theta, \Delta m^2)$ parameter space match very well those published by the MiniBooNE collaboration. We obtain similarly satisfactory results with the MiniBooNE antineutrino-mode data. With this agreement, we are confident we are capable of faithfully reproducing the data-analysis of MiniBooNE well enough to repeat the procedure for the decaying-sterile-neutrino hypothesis. 

\begin{figure}[ht]
\centering
\includegraphics[scale=0.35]{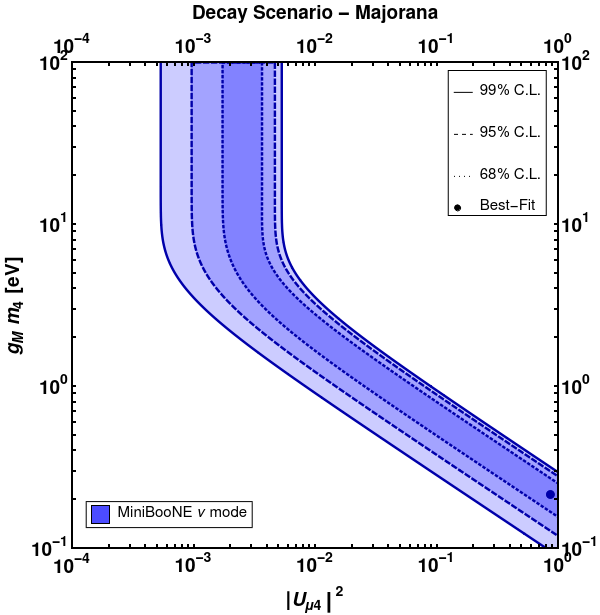}
\includegraphics[scale=0.35]{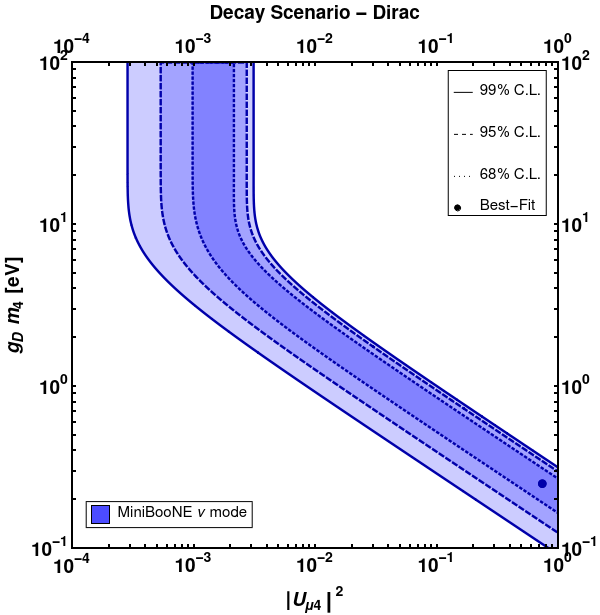}
\caption{Allowed regions of the $(|U_{\mu4}|^2,gm_4)$ parameter space when the decaying-sterile-neutrino hypothesis is matched against the MiniBooNE neutrino-mode data assuming Majorana (left) or Dirac neutrinos (right). The dots indicate the best-fit-point  and the lines represent the 99\% (solid), 95\% (dashed) and 68\% (dotted) C.L. curves.}
\label{fig:MB-nu-chisq}
\end{figure}

We generate neutrino event spectra for each set of decay parameters $\left( | U_{\mu 4} |^{2}, \ gm_{4} \right)$ and attempt to fit them to the MiniBooNE data, using a $\chi^{2}$-fit. The best-fit spectra to neutrino-mode data, in the case of Dirac and Majorana neutrinos are depicted, respectively, in black and blue in Fig.~\ref{fig:MB-spec}. 

\begin{figure}[h]
\centering
\includegraphics[scale=0.35]{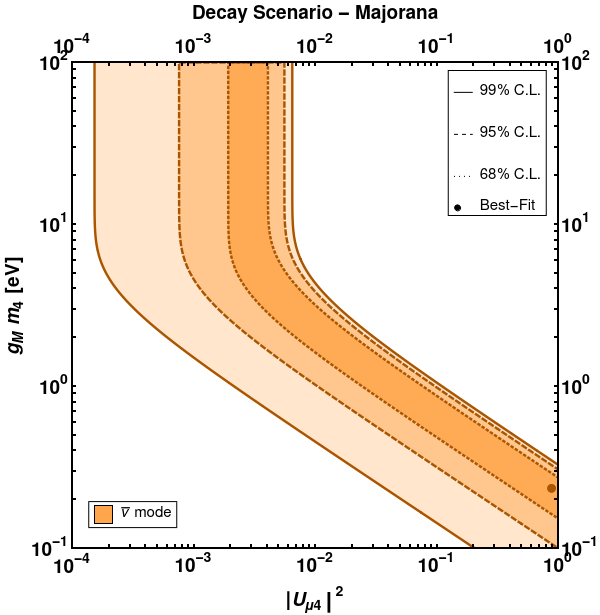}
\includegraphics[scale=0.35]{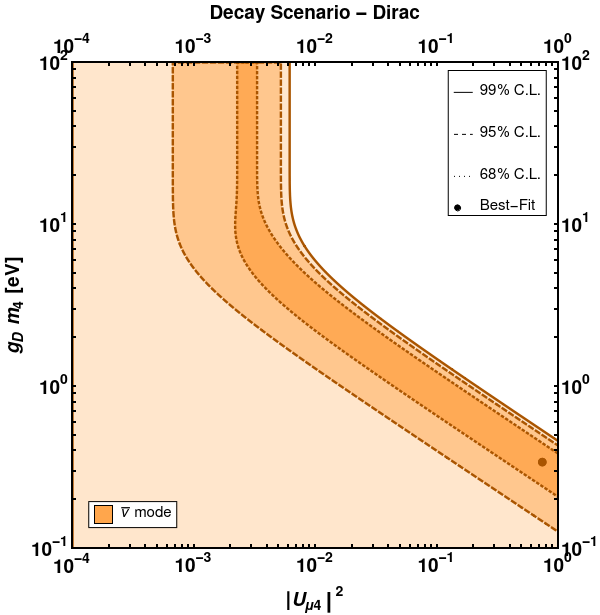}
\caption{Allowed regions of the $(|U_{\mu4}|^2,gm_4)$ parameter space when the decaying-sterile-neutrino hypothesis is matched against the MiniBooNE antineutrino-mode data assuming Majorana (left) or Dirac neutrinos (right). The dots indicate the best-fit-point and the lines represent the 99\% (solid), 95\% (dashed) and 68\% (dotted) C.L. curves}.
\label{fig:MB-anu-chisq}
\end{figure}

\begin{figure}[h]
\centering
\includegraphics[scale=0.35]{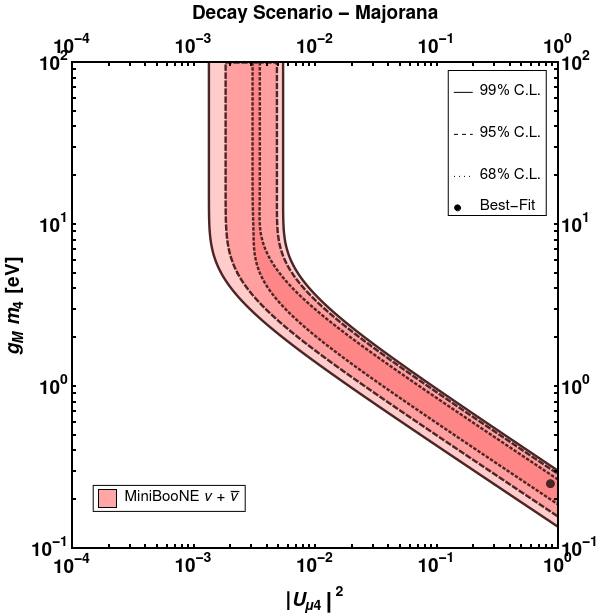}
\includegraphics[scale=0.35]{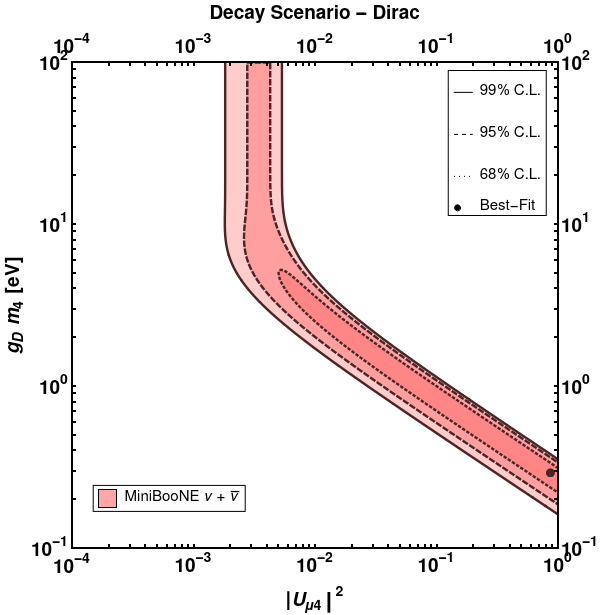}
\caption{Allowed regions of the $(|U_{\mu4}|^2,gm_4)$ parameter space when the decaying-sterile-neutrino hypothesis is matched against the combined MiniBooNE neutrino-mode and antineutrino-mode data assuming Majorana (left) or Dirac neutrinos (right). The dots indicate the best-fit-point  and the lines represent the 99\% (solid), 95\% (dashed) and 68\% (dotted) C.L. curves.}
\label{fig:MB-comb-chisq}
\end{figure}

For both neutrino-mode and antineutrino-mode data, the best fit point falls in the region where the decay is relatively slow. Hence, to zeroth order, a lower-energy $\nu_4$ decay more often than a higher-energy $\nu_4$. For the neutrino mode, we estimate the goodness-of-fit by comparing $\chi^2_{\rm min}$=11.08 (11.56) in the Dirac (Majorana) cases with nine degrees of freedom and conclude the fit is acceptable. For the antineutrino-mode, we estimate the goodness-of-fit by comparing $\chi^2_{\rm min}$=7.71 (6.66) in the Dirac (Majorana) cases with nine degrees of freedom and conclude the fit is also acceptable. The quality of these fits is similar to that of the oscillation fit. The allowed regions of the parameter space are depicted in Figs.~\ref{fig:MB-nu-chisq} (neutrino mode), \ref{fig:MB-anu-chisq} (antineutrino mode), and \ref{fig:MB-comb-chisq} (neutrino and antineutrino modes combined).


Unlike the LSND case, as advertised, the results of the two decay scenarios are similar for roughly similar values of $|U_{\mu4}|^2$. There is no obvious factor of two map between the Dirac and Majorana hypotheses, especially in the case of the antineutrino mode. This can be understood from the following. For the Majorana case, the channels which can in principle contribute to the observed event rates, for both neutrino and antineutrino runnings, are $\nu_{\mu} \rightarrow \nu_{e}$, $\nu_{\mu} \rightarrow \bar{\nu}_{e}$, $\bar{\nu}_{\mu} \rightarrow \nu_{e}$ and $\bar{\nu}_{\mu} \rightarrow \bar{\nu}_{e}$ (keeping in mind the facts that there is wrong-sign contamination\footnote{Presence of $\bar{\nu}_{\mu}$ in $\nu_{\mu}$-flux and $\nu_{\mu}$ in $\bar{\nu}_{\mu}$-flux.} in both the fluxes and that the MiniBooNE detector cannot distinguish an $e^{-}$ from an $e^{+}$). For the Dirac neutrinos, the helicity-flipping channels are irrelevant. In the case of neutrino-running, the wrong-sign contamination in the neutrino flux is tiny and therefore, there is negligible $\bar{\nu}_{\mu}\rightarrow \nu_{e}$ or $\bar{\nu}_{\mu}\rightarrow \bar{\nu}_{e}$ contribution to the event rates even if the transition probabilities in Eq.~(\ref{eq:pmue}) for the helicity-flipping channel is comparable to the helicity-conserving one. For the antineutrino running, all four channels are relevant as the wrong-sign contamination in the antineutrino fluxes  is rather large. In addition to the above arguments, one needs to take into account that the helicity-flipped daughter neutrinos peak softly; and the scattering cross-sections are different for neutrinos and antineutrinos. Thus, although $B_{e}$ for the Dirac case is twice that for the Majorana case; in the Majorana case, surplus decay channels and/or increased scattering cross-sections balance-out the situation and ultimately, we observe that similar values of the parameters yield similar-quality fits for the Majorana and Dirac hypothesis, especially in the case of antineutrino-mode data.

\subsection{LSND and MiniBooNE Combined}

\begin{figure}[h]
\centering
\includegraphics[scale=0.31]{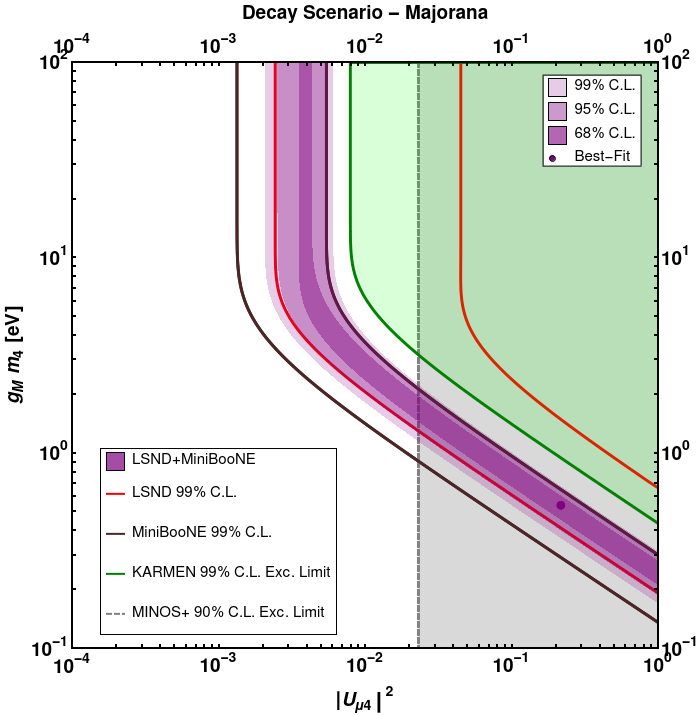}
\includegraphics[scale=0.31]{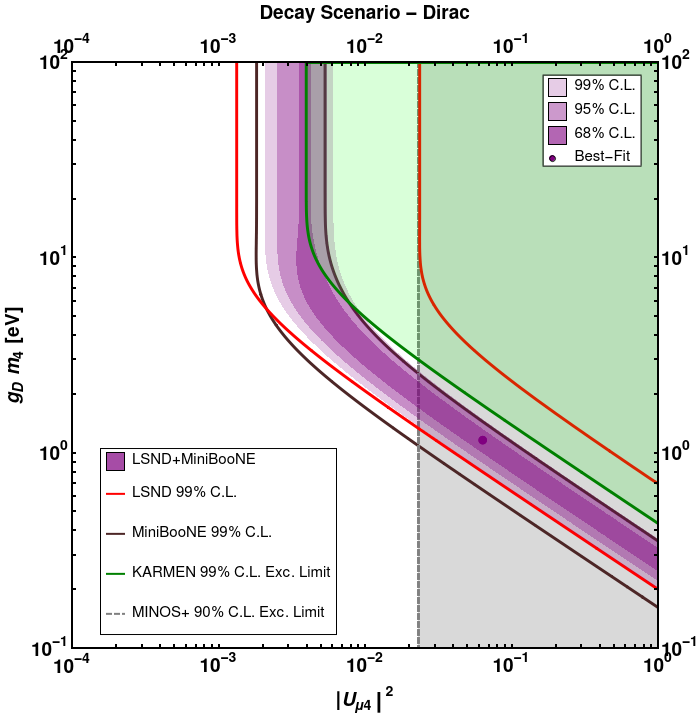}
\caption{Allowed regions at 99\% (lighter purple), 95\% (medium purple) and 68\% (darker purple) C.L. of the $(|U_{\mu4}|^2,gm_4)$ parameter space when the decaying-sterile-neutrino hypothesis is matched against the combined LSND data and MiniBooNE neutrino-mode and antineutrino-mode data assuming Majorana (left) or Dirac neutrinos (right). The dots indicate the best-fit-point. The region to the right of the vertical line is excluded by MINOS+ at the 90\% C.L.~\cite{Adamson:2017uda}. The green shaded region on the top-right of the green line is excluded by KARMEN at the 99\% C.L.}
\label{fig:LSND-MB-chisq}
\end{figure}

Next, we evaluate how well the decaying-sterile-neutrino hypothesis fits both LSND and MiniBooNE data by adding the $\chi^2$ obtained in the two independent analyses. The LSND-only  and MiniBooNE-only allowed regions of the parameter space are depicted in Fig.~\ref{fig:LSND-MB-chisq} to facilitate comparisons, along with the combined LSND+MiniBooNE allowed regions of the parameter space. The combined best-fit point, for the Dirac-neutrino scenario, is at $\left(|U_{\mu 4}|^2,\ g_{D}m_{4}\right) = \left(0.063,\ 1.17\ \rm{eV}\right)$ and $\chi^2_{\rm min}=45.33$. For 31 degrees of freedom (11+11+11-2), we estimate a p-value of several percent, which we deem to be reasonable. The event rates corresponding to the combined best-fit, for the Majorana-neutrino case are depicted in Figs.~\ref{fig:LSND-spec}, for LSND (gold color) and \ref{fig:MB-spec}, for MiniBooNE (neutrino-mode) (magenta). Note that the best-fit slightly undershoots the LSND data, and slightly overshoots those from MiniBooNE. The situation of the Majorana-neutrino scenario is similar; the quality of the fit is a little worse: $\chi^2_{\rm min}=48.34$.

\subsection{KARMEN, MINOS and MINOS+}\label{sub:minos}

The Karlsruhe Rutherford Medium Energy Neutrino (KARMEN) ran at the spallation neutrino source ISIS of the Rutherford Laboratory in the UK. We consider the data set corresponding to the experimental run from February 1997 to March 2001 \cite{Armbruster:2002mp}. The experiment impinges 800 MeV protons on a water-cooled $\rm{Ta-D_{2}}O$ target where $\pi^{+}$ per incident proton are produced. These $\pi^{+}$ are stopped completely and decay with a lifetime of $\tau_{\pi} = 26~\rm{ns}$ within the heavy target producing $\mu^{+}$ and $\nu_{\mu}$. The $\mu^{+}$ produced also decays at rest within the target with a lifetime $\tau_{\mu} = 2.2~\rm{\mu s}$ giving $e^{+},\nu_{e},\bar{\nu}_{\mu}$. Due to this large time separation the $\nu_{\mu}$ induced events can be cleanly separated from the $\bar{\nu}_{\mu}$ or $\nu_{e}$ induced events. The $\bar{\nu}_{\mu}$ and $\nu_{e}$ from the muon decay have a continuous spectra with the endpoint energy of $52.8~\rm{MeV}$. This data set corresponds to a total of $N_{\nu} = 2.71\times 10^{21}$ neutrinos for each flavor. The KARMEN detector consists of a liquid scintillation calorimeter situated at a mean distance of 17.7 m from the ISIS target and has a high energy resolution of $11.5\%/\sqrt{\rm{E\left(MeV\right)}}$. KARMEN observed a total of 15 inverse beta decay events compared against a background expectation of 15.8. Thus, it observed a null result for the $\bar{\nu}_{\mu}\rightarrow\bar{\nu}_{e}$ oscillations for $L/E_{\nu}\sim0.3-0.9~\rm{m/MeV}$. We follow the experimental details and analysis procedure described in 
Ref.~\cite{Armbruster:2002mp}, considering 9 energy bins between 16 MeV to 52 MeV and an overall normalization error of $10\%$ for both signal and backgrounds. We simulated the KARMEN experiment in GLoBES and performed fits to the oscillations and decay scenarios. For the case of oscillations, we get a $\chi^2_{\rm{min}}$ of 6.47 for 7 degrees of freedom and our result very closely resembles the results of Ref.~\cite{Armbruster:2002mp}. For the case of decay as well we we get a $\chi^2_{\rm min}$ of 6.47 for 7 degrees of freedom, for both Dirac and Majorana case; and our results are shown in Fig.~\ref{fig:LSND-MB-chisq}. Note that with this data set, we can only calculate constraints on the helicity-conserving $\bar{\nu}_{\mu}\rightarrow\bar{\nu}_{e}$ decay channel as the events due to the helicity flipping $\nu_{\mu}\rightarrow\bar{\nu}_{e}$ channel are not included in this sample due to a precise information regarding the timing of the events. 

MINOS \cite{Ambats:1998aa} is a long-baseline superbeam experiment based at Fermilab. The source of neutrinos is the NuMI beam facility at Fermilab \cite{Anderson:1998zza}. The experimental setup consists of a 1~kton near detector situated 1.04 km downstream and a 5.4~kton far detector situated 735~km away, on-axis in the Soudan underground laboratory. The primary goal of the MINOS experiment was to confirm, with an accelerator-based $\nu_{\mu}$-beam, the evidence for $\nu_{\mu}$-disappearance first seen in atmospheric experiments, measure the oscillation parameters $\sin^22\theta_{23}$ and $|\Delta m^2_{31}|$, and look for the subleading long-baseline $\nu_e$-appearance signal. For these purposes, MINOS looked at charged-current $\nu_{\mu}$-disappearance and $\nu_e$-appearance events in both neutrino and antineutrino modes \cite{Adamson:2014vgd}. It also measures neutral current events that are helpful in sterile-neutrino searches. Initially, MINOS operated with the low-energy tune of the NuMI beam that peaks at neutrino energies around $3~\rm{GeV}$. This was followed by running, referred to as MINOS+,  with the medium-energy tune of the NuMI beam, where the flux peaks at neutrino energies around $7~\rm{GeV}$. The most recent sterile neutrino searches were presented in \cite{Adamson:2017uda}. These results correspond to an exposure of $10.56\times10^{20}$~POT for MINOS and $5.80\times10^{20}$~POT for the MINOS+ experiment. Assuming the neutrino mass-eigenstates are stable, for $m_4\gg 10$~eV, the collaboration claims that the data constrain $|U_{\mu 4}|^2<2.3\times 10^{-2}$ at the 90\% C.L. Here, we take this result at face value and apply it to the decaying-sterile-neutrino scenarios of interest. 

Strictly speaking, the analysis presented in \cite{Adamson:2017uda} does not apply if the $\nu_4$ is unstable, for two reasons. One was already discussed. If one ignores the daughters of the neutrino decay, the $\nu_{\mu}$ survival probability depends on the $\nu_4$ lifetime, see Eq.~(\ref{eq:pmumu}). However, the difference between a stable and unstable $\nu_4$, as far as this contribution is concerned, is proportional to $|U_{\mu 4}|^4$, a factor $|U_{\mu 4}|^2$ smaller than the leading contribution. Since MINOS(+) is sensitive to $|U_{\mu 4}|^2$ values of order $10^{-2}$, the fact that $\nu_4$ can decay is irrelevant for this contribution to the disappearance analysis. The other potential impact of the decay is that the daughter $\nu_e$ of the $\nu_4$ decay can oscillate into a $\nu_{\mu}$ by the time it reaches the far detector. This extra contribution to the $\nu_{\mu}$ survival probability is, relative to the leading $|U_{\mu 4}|^2$-effect, suppressed by $|U_{e3}|^2\sim 0.02$ and hence very small. 

For the reasons discussed above, we take the constraint from the $\nu_{\mu}$ disappearance data to be $|U_{\mu 4}|^2<2.3\times 10^{-2}$ at the 90\% C.L. for all values of $gm_4$ of interest. This is represented by a vertical line in Fig.~\ref{fig:LSND-MB-chisq}. This constraint rules out the region of parameter corresponding to small $gm_4$ but leaves behind a healthy portion of the parameter space, including values of $gm_4$ small enough that the decay of $\nu_4$ is not necessarily prompt for the energies of interest. Since the Dirac hypothesis points to relatively smaller values of $|U_{\mu 4}|^2$, the allowed region of parameter space is ``larger'' in this case. 

One final note before proceeding. Given that, for large $gm_4$, we require $|U_{\mu4}|^2\lesssim 10^{-2}$ (and independent of $gm_4$), the bounds from meson leptonic decays on $g$ and $|U_{\mu4}|^2$, discussed in Sec.~\ref{sub:2}, translate into $gm_4\lesssim 10^{3}$~eV, saturated as $m_4$ approaches 1~MeV. 

Finally, we joined the null-disappearance results obtained by MINOS and KARMEN with the appearance results by LSND and MiniBooNE in one combined fit. The analysis was done by summing the $\chi^2$ functions of LSND, MiniBooNE and KARMEN and adding an penalty factor of $ \chi^2_{\rm penalty}=4.6\left(|U_{\mu 4}|^2/2.3 \times 10^{-2}\right)^2$ to describe the MINOS constraint. The combined LSND+MiniBooNE+KARMEN+MINOS allowed regions of the parameter space are shown in Fig.~\ref{fig:SBN-sens}. The combined best-fit point for Dirac case is at $\left(|U_{\mu 4}|^2,\ g_{D}m_{4}\right) = \left(0.0086,\ 3.41\ \rm{eV}\right)$ with $\chi^2_{\rm min}=56.42$ and for Majorana case is at $\left(|U_{\mu 4}|^2,\ g_{M}m_{4}\right) = \left(0.0086,\ 2.93\ \rm{eV}\right)$ with $\chi^2_{\rm min}=58.45$. Considering we have 40 degrees of freedom (11+11+11+9-2), we estimate a reasonable fit for both physics scenarios.

\subsection{SBN}\label{sub:sbn}

 The Short-Baseline Neutrino (SBN) Program is a set of three liquid argon detectors that will be aligned with the central axis of the BNB at Fermilab. Table~\ref{tab:detinfo} gives the SBN detector names, active masses and locations. According to the proposal~\cite{Antonello:2015lea}, the SBN Program is designed to address several anomalies in neutrino physics and will test, with the most sensitivity, the oscillation-interpretation to LSND and MiniBooNE data. 
     
     In order to explore the potential of the SBN Program to test the decaying-sterile neutrino model scenarios discussed here, we performed a sensitivity analysis considering only the neutrino-mode running for the BNB (see Section~\ref{sub:miniboone}). The generation of events as well as the $\chi^{2}$ function were implemented in GLoBES. The relevant details regarding the flux, the scattering cross-sections and efficiencies at the detectors are described in Ref.~\cite{Stenico:2018jpl} and the assumptions we make here are the same. We are considering only the $\nu_e$-appearance channel in order to estimate the sensitivity of the SBN Program. In the analysis, we imposed the same spectrum-normalization nuisance factor to the three detectors, since they receive neutrinos from the same source. The uncertainty related to the flux normalization was set to 15\%. Detailed descriptions of the signal and background for the appearance channel can also be found in Ref.~\cite{Stenico:2018jpl}.  
     
     
         \begin{table}[h]
\centering
\begin{adjustbox}{width=0.7\textwidth}
\begin{tabular}{||c c c||} 
 \hline
 \textbf{Detector} & \textbf{Active Mass} & \textbf{Distance from BNB target} \\ [0.5ex] 
 \hline\hline
SBND & 112 t & 110 m  \\ 
MicroBooNE & 89 t & 470 m  \\
ICARUS-T600 & 476 t & 600 m \\
 \hline\hline
\end{tabular}
 \end{adjustbox}
\caption{SBN detector active masses and distances from the local of the neutrino production.}
\label{tab:detinfo}
\end{table}

\begin{figure}[t]
\centering
\includegraphics[scale=0.35]{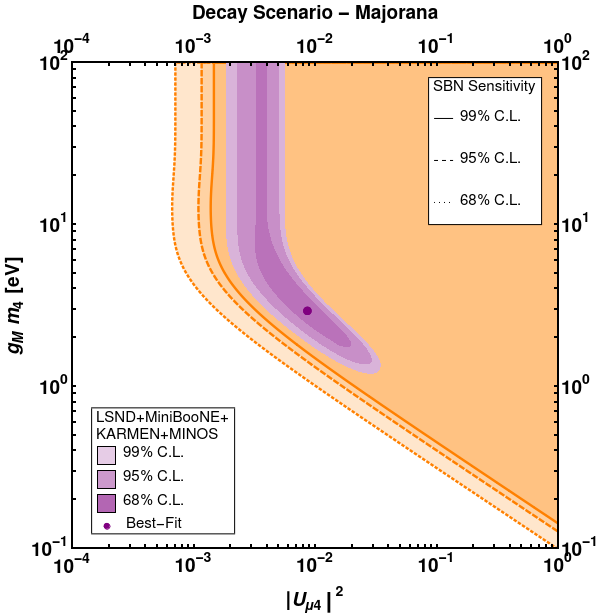}
\includegraphics[scale=0.35]{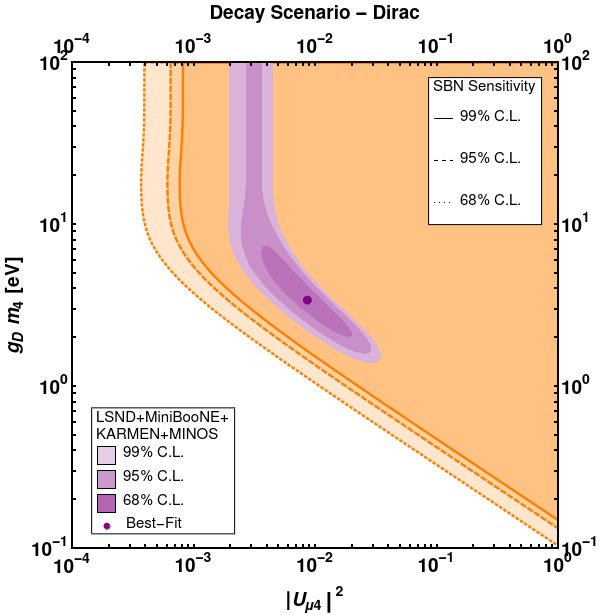}
\caption{Allowed regions at 99\% (lighter purple), 95\% (medium purple) and 68\% (darker purple) C.L. of the $(|U_{\mu4}|^2,gm_4)$ parameter space when the decaying-sterile-neutrino hypothesis is matched against the combined LSND, MiniBooNE and KARMEN data and MINOS constrains assuming Majorana (left) or Dirac neutrinos (right). The dots indicate the best-fit-point. In the same context, the orange regions indicate the sensitivity of the SBN Program at 99\% (solid line), 95\% (dashed line) and 68\% (dotted line) C.L. for Majorana (left) and Dirac neutrinos (right). We assume $6.6\times10^{20}$~POT for SBND and ICARUS and $1.32\times10^{21}$~POT for MicroBooNE.}
\label{fig:SBN-sens}
\end{figure}

The sensitivity of the SBN Program, assuming $6.6\times10^{20}$~POT for SBND and ICARUS (three nominal years of running) and $1.32\times10^{21}$~POT for MicroBooNE (six nominal years of running),  is depicted in Fig.~\ref{fig:SBN-sens}. The regions of the parameter space preferred by combined LSND and MiniBooNE are also depicted in order to facilitate comparisons. The SBN program can definitively test the decaying-sterile neutrino solution to the LSND and MiniBooNE data. 

\subsubsection{Sensitivity to non-zero neutrino decay effect on SBN}

\begin{figure}[t]
\centering
\includegraphics[scale=0.35]{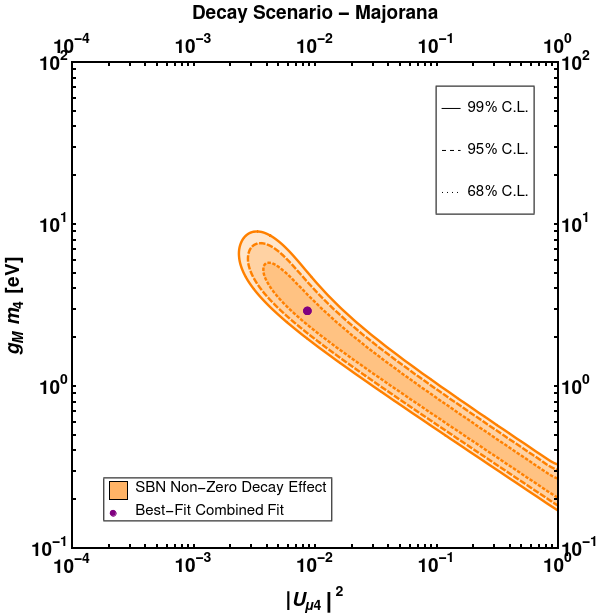}
\includegraphics[scale=0.35]{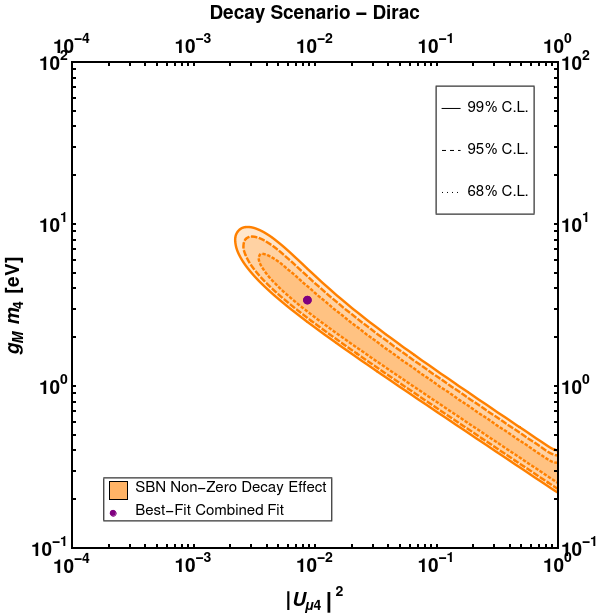}
\caption{SBN allowed regions for non-zero decay scenario parameters $(|U_{\mu4}|^2 ,gm_4)$ at 99\% (solid line), 95\% (dashed line) and 68\% (dotted line) C.L. for Majorana (left) and Dirac neutrinos (right). The dots indicate the best-fit-point from the LSND, MiniBooNE, KARMEN and MINOS combined analysis. In the same way of the SBN sensitivity analysis, we assume $6.6\times10^{20}$~POT for SBND and ICARUS and $1.32\times10^{21}$~POT for MicroBooNE.}
\label{fig:SBN-nonzero}
\end{figure}

Assuming the considered decaying-sterile neutrino model has a positive signal in SBN Program, we want to investigate now the capability of the experiment to measure the decay parameters $(|U_{\mu4}|^2 ,gm_4)$. To perform this analysis, we generated neutrino events in the same ``experimental'' configuration of SBN previous sensitivity analysis, but assuming now the data is given by non-zero values to $(|U_{\mu4}|^2 ,gm_4)$ parameters. For convenience, we will set the true values of the parameters at the correspondent best-fit points from LSND, MiniBooNE, KARMEN and MINOS combined analysis for Majorana and Dirac cases. The results we obtained are shown in Figure~\ref{fig:SBN-nonzero}: we have the allowed regions consistent with the computed events at the best-fit point for both Majorana (left panel) and Dirac (right panel) assumptions at $68.3\%$ of C. L. (dotted curve), $95\%$ of C. L. (dashed curve) and $99\%$ of C. L. (solid curve).

\section{Summary and conclusions} \label{sec:conc}

The excess of $\nu_e$- and $\bar{\nu}_e$-candidate events at MiniBooNE and LSND remains unexplained. The, arguably, simplest solution -- 3+1 neutrino-oscillation with a new mass-squared difference around 1~eV$^2$ -- is, however, severely constrained. If these data are indeed pointing to more new physics in the neutrino sector, it is likely that the new physics contains more ingredients than new neutrino mass-eigenstate that mix slightly with the active neutrinos. Here, we explored the hypothesis that there is a new neutrino mass-eigenstate $\nu_4$ and a new very light scalar particle $\phi$. $\nu_4$ and $\phi$ interact in such a way that $\nu_4\to\nu_e\phi$. Here, the excess of  $\nu_e$- and $\bar{\nu}_e$-candidate events at MiniBooNE and LSND are the daughter $\nu_e$ and $\bar{\nu}_e$ from $\nu_4$ and $\bar{\nu}_4$ decay. This hypothesis was first proposed in Ref.~\cite{PalomaresRuiz:2005vf} in order to address the LSND anomaly. 

We find a reasonable fit to the data of MiniBooNE and LSND, albeit the quality of the fit to only MiniBooNE and LSND data is not as good as the one obtained with the 3+1 neutrino-oscillations hypothesis. The decaying-sterile-neutrino hypothesis, however, can cleanly evade data from $\nu_{\mu}$-disappearance searches, which constrain $|U_{\mu4}|^2\lesssim 10^{-2}$, and is immune to searches involving $\nu_e$-disappearance.  We find that precision measurements of meson leptonic decays can also be satisfied as long as $1~{\rm MeV}\gtrsim m_4\gtrsim 10$~keV. The SBN program at Fermilab should be able to definitively test the decaying-sterile-neutrino hypothesis. We considered two different decay scenarios, one with Majorana neutrinos, one with Dirac neutrinos. The MiniBooNE and LSND data are such that both models fit the data with very similar efficacy. 

There are a few other new-physics solutions to the LSND and MiniBooNE data. Several, however, address only one data set or the other, including some recent, very interesting ideas~\cite{Magill:2018jla,Bertuzzo:2018itn,Ballett:2018ynz} that also postulate the existence of new light particles and new interactions. While the decaying-sterile-neutrino hypothesis explored here is not an excellent fit to both data sets -- especially the LSND data -- it seems to provide an interesting possibility. We hope the results presented here will inspire the collaborations -- they are the only ones capable of performing a proper fit to their data -- to investigate this possibility. 

We did not consider bounds from early-universe cosmology. The relatively large mixing between $\nu_s$ and $\nu_{\mu}$ indicates that it should be in thermal equilibrium in the early universe \cite{Hannestad:2015tea}. The fact that they decay quickly, however, should loosen bounds from, for example, big-bang nucleosynthesis. The new interaction between active and sterile neutrinos will also impact the dynamics of the early universe, and so will the new light degree of freedom $\phi$. More dynamics, including, for example, other couplings of $\phi$ to the active neutrinos, may help alleviate some of the potential tension. The exploration of these types of constraints is beyond the ambitions of this manuscript. 

Other manifestations of the sterile-neutrino decay hypothesis have been, very recently, discussed in the literature, including \cite{Fischer:2019fbw, Moulai:2019gpi, Dentler:2019dhz}. The work presented here share several similarities with these efforts but we explore, for the most part, a different region of the -- very large -- space of decaying-sterile-neutrino models. 

\begin{acknowledgments}
We thank Mona Dentler, Ivan Esteban, Joachim Kopp, and Pedro Machado  for sharing their analysis of the decaying-sterile-neutrino hypothesis before it became available on the arXivs.
G.V.S.~acknowledges useful discussions with Pedro Pasquini. 
The work of A.d.G.~was supported in part by DOE grant \#de-sc0010143.
 O.L.G.P.~is thankful for the support of FAPESP funding Grant  No. 2014/19164-6, FAEPEX funding grant 2391/2017 and 2541/2019,  CNPq grants 304715/2016-6 and 306565/2019-6. S.P.~thanks the support of the FAPESP funding grant No. 2017/02361-1. G.V.S.~is thankful for the support of FAPESP funding Grant  No. 2016/00272-9 and No. 2017/12904-2 and FAEPEX funding grant 2925/19. G.V.S. thanks the partial support of  FERMILAB-UNICAMP exchange agreement.
\end{acknowledgments}

\appendix
\section{Details on the MiniBooNE analysis}\label{app:a}

In this Section, we are going to describe the neutrino-only analysis, but the step works to antineutrino-only and combined analysis as well. In order to perform MiniBooNE analysis to decaying-sterile neutrino model, we generated an event spectrum correspondent to each set of parameters $(|U_{\mu4}|^2 ,gm_4)$ plotted in this work. After simulating the mentioned events, we analyse our ``pseudo'' data with the $\chi^2$ function defined by
   
\begin{equation}
\chi^2 = \sum_{i,j=1}^{N_e + N_\mu} (D_i - P_i)\mathcal{M}_{ij}^{-1}(D_j - P_j)
\end{equation}
        
\noindent where:

\begin{itemize}
\item $N_e$ is the number of the energy bins related to the observed electron neutrino CCQE events;
\item $N_{\mu}$ is the number of the energy bins related to the observed muon neutrino CCQE events;
\item $D_i$ is the element of a vector $D$ that contains $N_e$ + $N_\mu$ entries. The first $N_e$ entries correspond to the number of observed electron neutrino CCQE events in each of the $N_e$ energy bins. The followed $N_\mu$ entries correspond to the number of observed muon neutrino CCQE events in each of the $N_\mu$ energy bins;
\item $P_i$ is the element of a vector $P$ that contains $N_e$ entries of our predicted signal $S_i$ plus the estimated background $B_i$ for the electron neutrino events, followed by $N_\mu$ entries of the estimated muon neutrino events $M_i$ at MiniBooNE detector;
\item $\mathcal{M}_{ij}^{-1}$ is the inverse of the total $(N_e  + N_\mu)\times(N_e  + N_\mu)$ covariance matrix $\mathcal{M}_{ij}$, which includes all systematic and statistical uncertainties for the predicted events at vector $P$, and bin-to-bin systematic correlations.
\end{itemize}

\noindent The information about the number of the energy bins, the full content of the vector $D$, and the estimated electron neutrino background $B_i$ as well as muon neutrino CCQE events $M_i$ presented in vector $P$ were given by MiniBooNE collaboration at Ref.~\cite{mbdata1}. The covariance matrix $\mathcal{M}_{ij}$ must be obtained from vectors $D$ and $P$ and from the available \textit{fractional systematics-only covariance matrix} also given by the collaboration at Ref.~\cite{mbdata1}.

To derive $\mathcal{M}_{ij}$, we followed the step-by-step description available in Ref.~\cite{Karagiorgi:2010zz}. We are going to define the \textit{fractional systematics-only covariance matrix} as $\mathcal{M}_{kl}^{\rm frac}$. It consists of a $(N_e + N_e  + N_\mu)\times(N_e + N_e  + N_\mu)$ block matrix which has the form (full $\nu_\mu \rightarrow \nu_e$ conversion, $\nu_e$ BG, $\nu_\mu$), where

\begin{itemize}
    \item \textbf{full $\nu_\mu \rightarrow \nu_e$ conversion}: full $\nu_e$ transmutation events from $\nu_\mu$ flux. It consists of the initial $\nu_\mu$ a hundred percent converted in $\nu_e$ and then reconstructed and selected according to $\nu_e$ selection cuts;
    \item \textbf{$\nu_e$ BG}: estimated background $B_i$ for the electron neutrino events.;
    \item \textbf{$\nu_\mu$}: estimated muon neutrino CCQE events $M_i$.
\end{itemize}

First, we need to scale the matrix $\mathcal{M}_{kl}^{\rm frac}$ bin-by-bin to include the conversion probability correspondent to our signal. The resulting matrix ${M}_{kl}^{\rm sys}$ is given by:
              
\begin{equation}
\mathcal{M}_{kl}^{\rm sys} = \mathcal{M}_{kl}^{\rm frac} \cdot (P^{\prime}_k \cdot P^{\prime}_l),
\end{equation}

\noindent with $k,l = 1,... (N_e + N_e  + N_\mu)$. The vector $P^{\prime}$ contains $N_e$ entries of our signal events $S_i$, followed by $N_e$ entries of the estimated electron neutrino background $B_i$ and $ N_\mu$ entries of the estimated $\nu_\mu$ events $M_i$. Note that while $P^{\prime}$ has dimension $ (N_e + N_e + N_\mu)$, $P$ has dimension $ (N_e + N_\mu)$. 

The statistical error from our signal prediction is included by adding the elements $S_i$ to the diagonal elements of the $\mathcal{M}_{kl}^{\rm sys}$ for $k=1,...,N_e$:
        
\begin{equation}
\mathcal{M}_{kl}^{\rm{sys + stat}} = \mathcal{M}_{kl}^{\rm sys} + \delta_{kl} P^{\prime}_k
\end{equation}
        
\begin{figure}[t]
\centering
\includegraphics[scale=0.3]{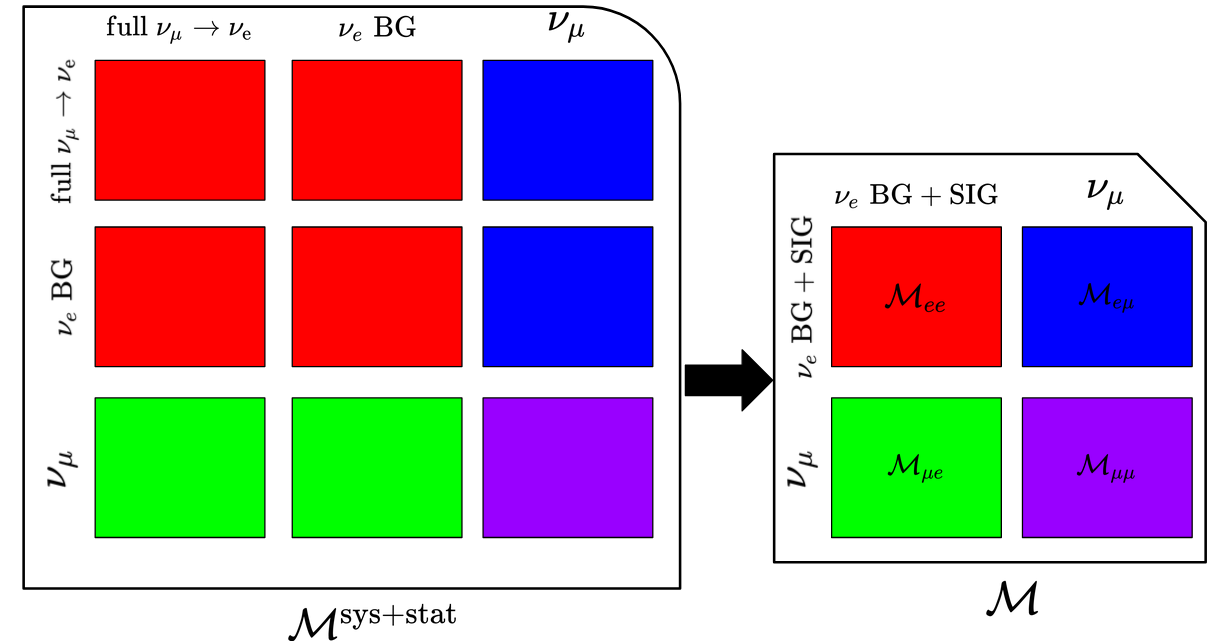}
\caption{Color scheme to collapse the matrix $\mathcal{M}^{\rm{sys + stat}}$ (left) into the matrix $\mathcal{M}$ (right) by overlapping blocks with the same color. Observe that the final matrix $\mathcal{M}$ is divided in the sub-blocks $\mathcal{M}_{e e}$, $\mathcal{M}_{\mu\mu}$, $\mathcal{M}_{e \mu}$ and $\mathcal{M}_{\mu e}$, which will be useful in the performance of electron neutrino appearance analysis.}
\label{fig:cov-matrix}
\end{figure}

Finally, we need to collapse the matrix $\mathcal{M}_{kl}^{\rm{sys + stat}}$ into $\mathcal{M}_{ij}$ and invert it to $\mathcal{M}_{ij}^{-1}$. In order to collapse $\mathcal{M}_{kl}^{\rm{sys + stat}}$, we follow the color pattern presented in Figure~\ref{fig:cov-matrix}, where we have  $\mathcal{M}^{\rm{sys + stat}}$ in the left and $\mathcal{M}$ in the right. Each block with the same color has the same dimension. The collapse of the matrix ${M}^{\rm{sys + stat}}$ means to overlap the blocks with the same color by \textit{summing} the elements with the correspondent positions among the blocks.
        
Once we obtained the correct covariance matrix to perform our analysis, we want to select the portion of the $\chi^2$ function that is related with the electron neutrino sample. The main reason for this is to study the impact of the decaying-sterile neutrino model in MiniBooNE appearance data, where the model has positive signal. For more details, see Section~\ref{sub:miniboone}. To do this, we follow the prescription in Appendix E.4 of Ref.~\cite{Kopp:2013vaa} and define the appearance $\chi_{\rm app}^2$ function as:
        
\begin{equation}
\chi_{\rm app}^2 = \chi^2 - C
\end{equation}
        
\noindent where $\chi^2$ contains all the information of matrix $\mathcal{M}$ and $C = (D_\mu - P_\mu) \mathcal{M}^{-1}_{\mu\mu} (D_\mu - P_\mu)$ includes only the systematic and statistical errors among muon neutrino events. The sub-block matrix $\mathcal{M}_{\mu\mu}$ is defined in Figure~\ref{fig:cov-matrix} (purple sub-block). The quantity $\chi_{\rm{app}}^2$ is what we consider as a final result to our analysis and removes the ``pure'' muon neutrino correlations, although is important to mention that correlation among electron an muon neutrino events is still taken into account in our MiniBooNE appearance analysis.
        
\bibliographystyle{JHEP}
\bibliography{sbl-extra}

\providecommand{\href}[2]{#2}\begingroup\raggedright\begin{thebibliography}{10}

\bibitem{Aguilar:2001ty}
{\scshape LSND} collaboration, A.~Aguilar-Arevalo et~al., \emph{{Evidence for
  neutrino oscillations from the observation of anti-neutrino(electron)
  appearance in a anti-neutrino(muon) beam}},
  \href{https://doi.org/10.1103/PhysRevD.64.112007}{\emph{Phys. Rev.}
  {\bfseries D64} (2001) 112007},
  [\href{https://arxiv.org/abs/hep-ex/0104049}{{\ttfamily hep-ex/0104049}}].

\bibitem{AguilarArevalo:2007it}
{\scshape MiniBooNE} collaboration, A.~A. Aguilar-Arevalo et~al., \emph{{A
  Search for electron neutrino appearance at the $\Delta m^{2} \sim
  1$~{eV}$^{2}$ scale}},
  \href{https://doi.org/10.1103/PhysRevLett.98.231801}{\emph{Phys. Rev. Lett.}
  {\bfseries 98} (2007) 231801},
  [\href{https://arxiv.org/abs/0704.1500}{{\ttfamily 0704.1500}}].

\bibitem{Aguilar-Arevalo:2018gpe}
{\scshape MiniBooNE} collaboration, A.~Aguilar-Arevalo et~al.,
  \emph{{Significant Excess of ElectronLike Events in the MiniBooNE
  Short-Baseline Neutrino Experiment}},
  \href{https://doi.org/10.1103/PhysRevLett.121.221801}{\emph{Phys. Rev. Lett.}
  {\bfseries 121} (2018) 221801},
  [\href{https://arxiv.org/abs/1805.12028}{{\ttfamily 1805.12028}}].

\bibitem{AguilarArevalo:2010wv}
{\scshape MiniBooNE} collaboration, A.~A. Aguilar-Arevalo et~al., \emph{{Event
  Excess in the MiniBooNE Search for $\bar \nu_\mu \rightarrow \bar \nu_e$
  Oscillations}},
  \href{https://doi.org/10.1103/PhysRevLett.105.181801}{\emph{Phys. Rev. Lett.}
  {\bfseries 105} (2010) 181801},
  [\href{https://arxiv.org/abs/1007.1150}{{\ttfamily 1007.1150}}].

\bibitem{Collin:2016aqd}
G.~H. Collin, C.~A. Arg{\"u}elles, J.~M. Conrad and M.~H. Shaevitz,
  \emph{{First Constraints on the Complete Neutrino Mixing Matrix with a
  Sterile Neutrino}},
  \href{https://doi.org/10.1103/PhysRevLett.117.221801}{\emph{Phys. Rev. Lett.}
  {\bfseries 117} (2016) 221801},
  [\href{https://arxiv.org/abs/1607.00011}{{\ttfamily 1607.00011}}].

\bibitem{Dentler:2018sju}
M.~Dentler, A.~Hern\'andez-Cabezudo, J.~Kopp, P.~A.~N. Machado, M.~Maltoni,
  I.~Martinez-Soler et~al., \emph{{Updated Global Analysis of Neutrino
  Oscillations in the Presence of eV-Scale Sterile Neutrinos}},
  \href{https://doi.org/10.1007/JHEP08(2018)010}{\emph{JHEP} {\bfseries 08}
  (2018) 010}, [\href{https://arxiv.org/abs/1803.10661}{{\ttfamily
  1803.10661}}].

\bibitem{Boser:2019rta}
S.~B{\"o}ser, C.~Buck, C.~Giunti, J.~Lesgourgues, L.~Ludhova, S.~Mertens
  et~al., \emph{{Status of Light Sterile Neutrino Searches}},
  \href{https://doi.org/10.1016/j.ppnp.2019.103736}{\emph{Prog. Part. Nucl.
  Phys.} {\bfseries 111} (2020) 103736},
  [\href{https://arxiv.org/abs/1906.01739}{{\ttfamily 1906.01739}}].

\bibitem{Diaz:2019fwt}
A.~Diaz, C.~A. Arg{\"u}elles, G.~H. Collin, J.~M. Conrad and M.~H. Shaevitz,
  \emph{{Where Are We With Light Sterile Neutrinos?}},
  \href{https://arxiv.org/abs/1906.00045}{{\ttfamily 1906.00045}}.

\bibitem{PalomaresRuiz:2005vf}
S.~Palomares-Ruiz, S.~Pascoli and T.~Schwetz, \emph{{Explaining LSND by a
  decaying sterile neutrino}},
  \href{https://doi.org/10.1088/1126-6708/2005/09/048}{\emph{JHEP} {\bfseries
  09} (2005) 048}, [\href{https://arxiv.org/abs/hep-ph/0505216}{{\ttfamily
  hep-ph/0505216}}].

\bibitem{Gninenko:2009ks}
S.~N. Gninenko, \emph{{The MiniBooNE anomaly and heavy neutrino decay}},
  \href{https://doi.org/10.1103/PhysRevLett.103.241802}{\emph{Phys. Rev. Lett.}
  {\bfseries 103} (2009) 241802},
  [\href{https://arxiv.org/abs/0902.3802}{{\ttfamily 0902.3802}}].

\bibitem{Gninenko:2009yf}
S.~N. Gninenko and D.~S. Gorbunov, \emph{{The MiniBooNE anomaly, the decay
  $D^{+}_{s} \to \mu^{+}\nu_{\mu}$ and heavy sterile neutrino}},
  \href{https://doi.org/10.1103/PhysRevD.81.075013}{\emph{Phys. Rev.}
  {\bfseries D81} (2010) 075013},
  [\href{https://arxiv.org/abs/0907.4666}{{\ttfamily 0907.4666}}].

\bibitem{Gninenko:2010pr}
S.~N. Gninenko, \emph{{A resolution of puzzles from the LSND, KARMEN, and
  MiniBooNE experiments}},
  \href{https://doi.org/10.1103/PhysRevD.83.015015}{\emph{Phys. Rev.}
  {\bfseries D83} (2011) 015015},
  [\href{https://arxiv.org/abs/1009.5536}{{\ttfamily 1009.5536}}].

\bibitem{Dib:2011jh}
C.~Dib, J.~C. Helo, S.~Kovalenko and I.~Schmidt, \emph{{Sterile neutrino decay
  explanation of LSND and MiniBooNE anomalies}},
  \href{https://doi.org/10.1103/PhysRevD.84.071301}{\emph{Phys. Rev.}
  {\bfseries D84} (2011) 071301},
  [\href{https://arxiv.org/abs/1105.4664}{{\ttfamily 1105.4664}}].

\bibitem{Schechter:1981cv}
J.~Schechter and J.~W.~F. Valle, \emph{{Neutrino Decay and Spontaneous
  Violation of Lepton Number}},
  \href{https://doi.org/10.1103/PhysRevD.25.774}{\emph{Phys. Rev.} {\bfseries
  D25} (1982) 774}.

\bibitem{PhysRevLett.45.1926}
Y.~Chikashige, R.~N. Mohapatra and R.~D. Peccei, \emph{Spontaneously broken
  lepton number and cosmological constraints on the neutrino mass spectrum},
  \href{https://doi.org/10.1103/PhysRevLett.45.1926}{\emph{Phys. Rev. Lett.}
  {\bfseries 45} (Dec, 1980) 1926--1929}.

\bibitem{Berezhiani:1990sy}
Z.~G. Berezhiani and M.~{\relax Yu}. Khlopov, \emph{{Physics of cosmological
  dark matter in the theory of broken family symmetry. (In Russian)}},
  {\emph{Sov. J. Nucl. Phys.} {\bfseries 52} (1990) 60--64}.

\bibitem{1988SvA....32..127D}
A.~G. {Doroshkevich}, A.~A. {Klypin} and M.~Y. {Khlopov}, \emph{{Cosmological
  Models with Unstable Neutrinos}}, {\emph{Sov. Astron.} {\bfseries 32} (Apr.,
  1988) 127}.

\bibitem{GELMINI1981411}
G.~Gelmini and M.~Roncadelli, \emph{Left-handed neutrino mass scale and
  spontaneously broken lepton number},
  \href{https://doi.org/https://doi.org/10.1016/0370-2693(81)90559-1}{\emph{Physics
  Letters B} {\bfseries 99} (1981) 411 -- 415}.

\bibitem{Gelmini:1983ea}
G.~B. Gelmini and J.~W.~F. Valle, \emph{{Fast Invisible Neutrino Decays}},
  \href{https://doi.org/10.1016/0370-2693(84)91258-9}{\emph{Phys. Lett.}
  {\bfseries 142B} (1984) 181--187}.

\bibitem{BERTOLINI1988714}
S.~Bertolini and A.~Santamaria, \emph{The doublet majoron model and solar
  neutrino oscillations},
  \href{https://doi.org/https://doi.org/10.1016/0550-3213(88)90100-9}{\emph{Nuclear
  Physics B} {\bfseries 310} (1988) 714 -- 742}.

\bibitem{SANTAMARIA1987423}
A.~Santamaria and J.~Valle, \emph{Spontaneous r parity violation in
  supersymmetry: A model for solar neutrino oscillations},
  \href{https://doi.org/https://doi.org/10.1016/0370-2693(87)90042-6}{\emph{Physics
  Letters B} {\bfseries 195} (1987) 423 -- 428}.

\bibitem{Cogollo:2008zc}
D.~Cogollo, H.~Diniz, C.~A. de~S.~Pires and P.~S. Rodrigues~da Silva,
  \emph{{The Seesaw mechanism at TeV scale in the 3-3-1 model with right-handed
  neutrinos}}, \href{https://doi.org/10.1140/epjc/s10052-008-0749-5}{\emph{Eur.
  Phys. J.} {\bfseries C58} (2008) 455--461},
  [\href{https://arxiv.org/abs/0806.3087}{{\ttfamily 0806.3087}}].

\bibitem{deGouvea:2019goq}
A.~de~Gouvêa, I.~Martinez-Soler and M.~Sen, \emph{{Impact of neutrino decays
  on the supernova neutronization-burst flux}},
  \href{https://doi.org/10.1103/PhysRevD.101.043013}{\emph{Phys. Rev. D}
  {\bfseries 101} (2020) 043013},
  [\href{https://arxiv.org/abs/1910.01127}{{\ttfamily 1910.01127}}].

\bibitem{Lindner:2001fx}
M.~Lindner, T.~Ohlsson and W.~Winter, \emph{{A Combined treatment of neutrino
  decay and neutrino oscillations}},
  \href{https://doi.org/10.1016/S0550-3213(01)00237-1}{\emph{Nucl. Phys.}
  {\bfseries B607} (2001) 326--354},
  [\href{https://arxiv.org/abs/hep-ph/0103170}{{\ttfamily hep-ph/0103170}}].

\bibitem{Kim:1990km}
C.~W. Kim and W.~P. Lam, \emph{{Some remarks on neutrino decay via a
  Nambu-Goldstone boson}},
  \href{https://doi.org/10.1142/S0217732390000354}{\emph{Mod. Phys. Lett.}
  {\bfseries A5} (1990) 297--299}.

\bibitem{Peres:2000ic}
O.~L.~G. Peres and A.~{\relax Yu}. Smirnov, \emph{{(3+1) spectrum of neutrino
  masses: A Chance for LSND?}},
  \href{https://doi.org/10.1016/S0550-3213(01)00012-8}{\emph{Nucl. Phys.}
  {\bfseries B599} (2001) 3},
  [\href{https://arxiv.org/abs/hep-ph/0011054}{{\ttfamily hep-ph/0011054}}].

\bibitem{deGouvea:2015euy}
A.~de~Gouv\^ea and A.~Kobach, \emph{{Global Constraints on a Heavy Neutrino}},
  \href{https://doi.org/10.1103/PhysRevD.93.033005}{\emph{Phys. Rev.}
  {\bfseries D93} (2016) 033005},
  [\href{https://arxiv.org/abs/1511.00683}{{\ttfamily 1511.00683}}].

\bibitem{Bryman:2019bjg}
D.~A. Bryman and R.~Shrock, \emph{{Constraints on Sterile Neutrinos in the MeV
  to GeV Mass Range}},
  \href{https://doi.org/10.1103/PhysRevD.100.073011}{\emph{Phys. Rev.}
  {\bfseries D100} (2019) 073011},
  [\href{https://arxiv.org/abs/1909.11198}{{\ttfamily 1909.11198}}].

\bibitem{Pasquini:2015fjv}
P.~S. Pasquini and O.~L.~G. Peres, \emph{{Bounds on Neutrino-Scalar Yukawa
  Coupling}}, \href{https://doi.org/10.1103/PhysRevD.93.053007,
  10.1103/PhysRevD.93.079902}{\emph{Phys. Rev.} {\bfseries D93} (2016) 053007},
  [\href{https://arxiv.org/abs/arXiv:1511.01811}{{\ttfamily
  arXiv:1511.01811}}].

\bibitem{Athanassopoulos:1996ds}
{\scshape LSND} collaboration, C.~Athanassopoulos et~al., \emph{{The Liquid
  scintillator neutrino detector and LAMPF neutrino source}},
  \href{https://doi.org/10.1016/S0168-9002(96)01155-2}{\emph{Nucl. Instrum.
  Meth. A} {\bfseries 388} (1997) 149--172},
  [\href{https://arxiv.org/abs/nucl-ex/9605002}{{\ttfamily nucl-ex/9605002}}].

\bibitem{Conrad:2016sve}
J.~M. Conrad and M.~H. Shaevitz, \emph{{Sterile Neutrinos: An Introduction to
  Experiments}}, \href{https://doi.org/10.1142/9789813226098_0010}{\emph{Adv.
  Ser. Direct. High Energy Phys.} {\bfseries 28} (2018) 391--442},
  [\href{https://arxiv.org/abs/1609.07803}{{\ttfamily 1609.07803}}].

\bibitem{Huber:2004ka}
P.~Huber, M.~Lindner and W.~Winter, \emph{{Simulation of long-baseline neutrino
  oscillation experiments with GLoBES (General Long Baseline Experiment
  Simulator)}}, \href{https://doi.org/10.1016/j.cpc.2005.01.003}{\emph{Comput.
  Phys. Commun.} {\bfseries 167} (2005) 195},
  [\href{https://arxiv.org/abs/hep-ph/0407333}{{\ttfamily hep-ph/0407333}}].

\bibitem{Huber:2007ji}
P.~Huber, J.~Kopp, M.~Lindner, M.~Rolinec and W.~Winter, \emph{{New features in
  the simulation of neutrino oscillation experiments with GLoBES 3.0: General
  Long Baseline Experiment Simulator}},
  \href{https://doi.org/10.1016/j.cpc.2007.05.004}{\emph{Comput. Phys. Commun.}
  {\bfseries 177} (2007) 432--438},
  [\href{https://arxiv.org/abs/hep-ph/0701187}{{\ttfamily hep-ph/0701187}}].

\bibitem{Strumia:2003zx}
A.~Strumia and F.~Vissani, \emph{{Precise quasielastic neutrino/nucleon
  cross-section}},
  \href{https://doi.org/10.1016/S0370-2693(03)00616-6}{\emph{Phys. Lett.}
  {\bfseries B564} (2003) 42--54},
  [\href{https://arxiv.org/abs/astro-ph/0302055}{{\ttfamily
  astro-ph/0302055}}].

\bibitem{AguilarArevalo:2008qa}
{\scshape MiniBooNE} collaboration, A.~A. Aguilar-Arevalo et~al., \emph{{The
  MiniBooNE Detector}},
  \href{https://doi.org/10.1016/j.nima.2008.10.028}{\emph{Nucl. Instrum. Meth.}
  {\bfseries A599} (2009) 28--46},
  [\href{https://arxiv.org/abs/0806.4201}{{\ttfamily 0806.4201}}].

\bibitem{mbdata1}
MiniBooNE, ``Data release for 1805.12028.''
  \url{https://www-boone.fnal.gov/for_physicists/data_release/nue2018/}.

\bibitem{AguilarArevalo:2008yp}
{\scshape MiniBooNE} collaboration, A.~A. Aguilar-Arevalo et~al., \emph{{The
  Neutrino Flux prediction at MiniBooNE}},
  \href{https://doi.org/10.1103/PhysRevD.79.072002}{\emph{Phys. Rev.}
  {\bfseries D79} (2009) 072002},
  [\href{https://arxiv.org/abs/0806.1449}{{\ttfamily 0806.1449}}].

\bibitem{mbdata2}
MiniBooNE, ``Data release for 1207.4809.''
  \url{https://www-boone.fnal.gov/for_physicists/data_release/nue_nuebar_2012/efficiency/MB_nu_nubar_combined_release.html}.

\bibitem{Kopp:2013vaa}
J.~Kopp, P.~A.~N. Machado, M.~Maltoni and T.~Schwetz, \emph{{Sterile Neutrino
  Oscillations: The Global Picture}},
  \href{https://doi.org/10.1007/JHEP05(2013)050}{\emph{JHEP} {\bfseries 05}
  (2013) 050}, [\href{https://arxiv.org/abs/1303.3011}{{\ttfamily 1303.3011}}].

\bibitem{Adamson:2017uda}
{\scshape MINOS+} collaboration, P.~Adamson et~al., \emph{{Search for sterile
  neutrinos in MINOS and MINOS+ using a two-detector fit}},
  \href{https://doi.org/10.1103/PhysRevLett.122.091803}{\emph{Phys. Rev. Lett.}
  {\bfseries 122} (2019) 091803},
  [\href{https://arxiv.org/abs/1710.06488}{{\ttfamily 1710.06488}}].

\bibitem{Armbruster:2002mp}
{\scshape KARMEN} collaboration, B.~Armbruster et~al., \emph{{Upper limits for
  neutrino oscillations $\bar{\nu}_{\mu} \to \bar{\nu}_e$ from muon decay at
  rest}}, \href{https://doi.org/10.1103/PhysRevD.65.112001}{\emph{Phys. Rev. D}
  {\bfseries 65} (2002) 112001},
  [\href{https://arxiv.org/abs/hep-ex/0203021}{{\ttfamily hep-ex/0203021}}].

\bibitem{Ambats:1998aa}
{\scshape MINOS} collaboration, I.~Ambats et~al., ``{The MINOS Detectors
  Technical Design Report}.''
  \url{http://lss.fnal.gov/archive/design/fermilab-design-1998-02.pdf}, 1998.

\bibitem{Anderson:1998zza}
K.~Anderson et~al., ``{The NuMI Facility Technical Design Report}.''
  \url{http://lss.fnal.gov/cgi-bin/find_paper.pl?design-1998-01.pdf}, 1998.
\newblock 10.2172/1156372.

\bibitem{Adamson:2014vgd}
{\scshape MINOS} collaboration, P.~Adamson et~al., \emph{{Combined analysis of
  $\nu_{\mu}$ disappearance and $\nu_{\mu} \rightarrow \nu_{e}$ appearance in
  MINOS using accelerator and atmospheric neutrinos}},
  \href{https://doi.org/10.1103/PhysRevLett.112.191801}{\emph{Phys. Rev. Lett.}
  {\bfseries 112} (2014) 191801},
  [\href{https://arxiv.org/abs/1403.0867}{{\ttfamily 1403.0867}}].

\bibitem{Antonello:2015lea}
{\scshape MicroBooNE, LAr1-ND, ICARUS-WA104} collaboration, M.~Antonello
  et~al., \emph{{A Proposal for a Three Detector Short-Baseline Neutrino
  Oscillation Program in the Fermilab Booster Neutrino Beam}},
  \href{https://arxiv.org/abs/1503.01520}{{\ttfamily 1503.01520}}.

\bibitem{Stenico:2018jpl}
G.~V. Stenico, D.~V. Forero and O.~L.~G. Peres, \emph{{A Short Travel for
  Neutrinos in Large Extra Dimensions}},
  \href{https://doi.org/10.1007/JHEP11(2018)155}{\emph{JHEP} {\bfseries 11}
  (2018) 155}, [\href{https://arxiv.org/abs/1808.05450}{{\ttfamily
  1808.05450}}].

\bibitem{Magill:2018jla}
G.~Magill, R.~Plestid, M.~Pospelov and Y.-D. Tsai, \emph{{Dipole Portal to
  Heavy Neutral Leptons}},
  \href{https://doi.org/10.1103/PhysRevD.98.115015}{\emph{Phys. Rev.}
  {\bfseries D98} (2018) 115015},
  [\href{https://arxiv.org/abs/1803.03262}{{\ttfamily 1803.03262}}].

\bibitem{Bertuzzo:2018itn}
E.~Bertuzzo, S.~Jana, P.~A.~N. Machado and R.~Zukanovich~Funchal, \emph{{Dark
  Neutrino Portal to Explain MiniBooNE excess}},
  \href{https://doi.org/10.1103/PhysRevLett.121.241801}{\emph{Phys. Rev. Lett.}
  {\bfseries 121} (2018) 241801},
  [\href{https://arxiv.org/abs/1807.09877}{{\ttfamily 1807.09877}}].

\bibitem{Ballett:2018ynz}
P.~Ballett, S.~Pascoli and M.~Ross-Lonergan, \emph{{U(1)' mediated decays of
  heavy sterile neutrinos in MiniBooNE}},
  \href{https://doi.org/10.1103/PhysRevD.99.071701}{\emph{Phys. Rev.}
  {\bfseries D99} (2019) 071701},
  [\href{https://arxiv.org/abs/1808.02915}{{\ttfamily 1808.02915}}].

\bibitem{Hannestad:2015tea}
S.~Hannestad, R.~S. Hansen, T.~Tram and Y.~Y.~Y. Wong, \emph{{Active-sterile
  neutrino oscillations in the early Universe with full collision terms}},
  \href{https://doi.org/10.1088/1475-7516/2015/08/019}{\emph{JCAP} {\bfseries
  1508} (2015) 019}, [\href{https://arxiv.org/abs/1506.05266}{{\ttfamily
  1506.05266}}].

\bibitem{Fischer:2019fbw}
O.~Fischer, A.~Hern\'andez-Cabezudo and T.~Schwetz, \emph{{Explaining the
  MiniBooNE excess by a decaying sterile neutrino with mass in the 250 MeV
  range}}, \href{https://doi.org/10.1103/PhysRevD.101.075045}{\emph{Phys. Rev.
  D} {\bfseries 101} (2020) 075045},
  [\href{https://arxiv.org/abs/1909.09561}{{\ttfamily 1909.09561}}].

\bibitem{Moulai:2019gpi}
M.~Moulai, C.~Argüelles, G.~Collin, J.~Conrad, A.~Diaz and M.~Shaevitz,
  \emph{{Combining Sterile Neutrino Fits to Short Baseline Data with IceCube
  Data}}, \href{https://doi.org/10.1103/PhysRevD.101.055020}{\emph{Phys. Rev.
  D} {\bfseries 101} (2020) 055020},
  [\href{https://arxiv.org/abs/1910.13456}{{\ttfamily 1910.13456}}].

\bibitem{Dentler:2019dhz}
M.~Dentler, I.~Esteban, J.~Kopp and P.~Machado, \emph{{Decaying Sterile
  Neutrinos and the Short Baseline Oscillation Anomalies}},
  \href{https://arxiv.org/abs/1911.01427}{{\ttfamily 1911.01427}}.

\bibitem{Karagiorgi:2010zz}
G.~S. Karagiorgi, ``{Searches for New Physics at MiniBooNE: Sterile Neutrinos
  and Mixing Freedom}.''
  \href{https://www.osti.gov/biblio/1000269}{https://www.osti.gov/biblio/1000269},
  2010.
\newblock 10.2172/1000269.

\end{thebibliography}\endgroup

\end{document}